\documentclass[ALICE,manyauthors]{cernphprep}
\usepackage[comma,square,numbers,sort&compress]{natbib}
\usepackage{hyperref}
\usepackage{lineno}
\usepackage{xspace}
  \usepackage[separate-uncertainty=true,locale=US]{siunitx}
  \DeclareSIUnit[quantity-product = ]\percent{\char`\%}
\usepackage{physics}
\usepackage{booktabs}
\usepackage{adjustbox}
\usepackage{rotating}
\usepackage[T1]{fontenc}
\usepackage{orcidlink}

\begin{document}
%

\newcommand{\pp}           {pp\xspace}
\newcommand{\ppbar}        {\mbox{$\mathrm {p\overline{p}}$}\xspace}
\newcommand{\XeXe}         {\mbox{Xe--Xe}\xspace}
\newcommand{\PbPb}         {\mbox{Pb--Pb}\xspace}
\newcommand{\pPb}          {\mbox{p--Pb}\xspace}
\newcommand{\AuAu}         {\mbox{Au--Au}\xspace}
\newcommand{\dAu}          {\mbox{d--Au}\xspace}

\newcommand{\snn}          {\ensuremath{\sqrt{s_{\mathrm{NN}}}}\xspace}
\newcommand{\pt}           {\ensuremath{p_{\rm T}}\xspace}
\newcommand{\mt}           {\ensuremath{m_{\rm T}}\xspace}
\newcommand{\meanpt}       {$\langle p_{\mathrm{T}}\rangle$\xspace}
\newcommand{\ycms}         {\ensuremath{y_{\rm CMS}}\xspace}
\newcommand{\ylab}         {\ensuremath{y_{\rm lab}}\xspace}
\newcommand{\etarange}[1]  {\mbox{$\left | \eta \right |~<~#1$}}
\newcommand{\yrange}[1]    {\mbox{$\left | y \right |~<~#1$}}
\newcommand{\dndy}         {\ensuremath{\mathrm{d}N_\mathrm{ch}/\mathrm{d}y}\xspace}
\newcommand{\dndeta}       {\ensuremath{\mathrm{d}N_\mathrm{ch}/\mathrm{d}\eta}\xspace}
\newcommand{\avdndeta}     {\ensuremath{\langle\dndeta\rangle}\xspace}
\newcommand{\dNdy}         {\ensuremath{\mathrm{d}N_\mathrm{ch}/\mathrm{d}y}\xspace}
\newcommand{\Npart}        {\ensuremath{N_\mathrm{part}}\xspace}
\newcommand{\Ncoll}        {\ensuremath{N_\mathrm{coll}}\xspace}
\newcommand{\dEdx}         {\ensuremath{\textrm{d}E/\textrm{d}x}\xspace}
\newcommand{\RpPb}         {\ensuremath{R_{\rm pPb}}\xspace}

\newcommand{\nineH}        {$\sqrt{s}~=~0.9$~Te\kern-.1emV\xspace}
\newcommand{\seven}        {$\sqrt{s}~=~7$~Te\kern-.1emV\xspace}
\newcommand{\twoH}         {$\sqrt{s}~=~0.2$~Te\kern-.1emV\xspace}
\newcommand{\twosevensix}  {$\sqrt{s}~=~2.76$~Te\kern-.1emV\xspace}
\newcommand{\five}         {$\sqrt{s}~=~5.02$~Te\kern-.1emV\xspace}
\newcommand{\twosevensixnn}{$\sqrt{s_{\mathrm{NN}}}~=~2.76$~Te\kern-.1emV\xspace}
\newcommand{\fivenn}       {$\sqrt{s_{\mathrm{NN}}}~=~5.02$~Te\kern-.1emV\xspace}
\newcommand{\LT}           {L{\'e}vy-Tsallis\xspace}
\newcommand{\GeVc}         {Ge\kern-.1emV/$c$\xspace}
\newcommand{\MeVc}         {Me\kern-.1emV/$c$\xspace}
\newcommand{\GeVmass}      {Ge\kern-.2emV/$c^2$\xspace}
\newcommand{\MeVmass}      {Me\kern-.2emV/$c^2$\xspace}
\newcommand{\lumi}         {\ensuremath{\mathcal{L}}\xspace}

\newcommand{\ITS}          {\rm{ITS}\xspace}
\newcommand{\TOF}          {\rm{TOF}\xspace}
\newcommand{\ZDC}          {\rm{ZDC}\xspace}
\newcommand{\ZDCs}         {\rm{ZDCs}\xspace}
\newcommand{\ZNA}          {\rm{ZNA}\xspace}
\newcommand{\ZNC}          {\rm{ZNC}\xspace}
\newcommand{\SPD}          {\rm{SPD}\xspace}
\newcommand{\SDD}          {\rm{SDD}\xspace}
\newcommand{\SSD}          {\rm{SSD}\xspace}
\newcommand{\TPC}          {\rm{TPC}\xspace}
\newcommand{\TRD}          {\rm{TRD}\xspace}
\newcommand{\VZERO}        {\rm{V0}\xspace}
\newcommand{\VZEROA}       {\rm{V0A}\xspace}
\newcommand{\VZEROC}       {\rm{V0C}\xspace}
\newcommand{\Vdecay} 	   {\ensuremath{V^{0}}\xspace}

\newcommand{\ee}           {\ensuremath{e^{+}e^{-}}} 
\newcommand{\pip}          {\ensuremath{\pi^{+}}\xspace}
\newcommand{\pim}          {\ensuremath{\pi^{-}}\xspace}
\newcommand{\kap}          {\ensuremath{\rm{K}^{+}}\xspace}
\newcommand{\kam}          {\ensuremath{\rm{K}^{-}}\xspace}
\newcommand{\pbar}         {\ensuremath{\rm\overline{p}}\xspace}
\newcommand{\kzero}        {\ensuremath{{\rm K}^{0}_{\rm{S}}}\xspace}
\newcommand{\lmb}          {\ensuremath{\Lambda}\xspace}
\newcommand{\almb}         {\ensuremath{\overline{\Lambda}}\xspace}
\newcommand{\Om}           {\ensuremath{\Omega^-}\xspace}
\newcommand{\Mo}           {\ensuremath{\overline{\Omega}^+}\xspace}
\newcommand{\X}            {\ensuremath{\Xi^-}\xspace}
\newcommand{\Ix}           {\ensuremath{\overline{\Xi}^+}\xspace}
\newcommand{\Xis}          {\ensuremath{\Xi^{\pm}}\xspace}
\newcommand{\Oms}          {\ensuremath{\Omega^{\pm}}\xspace}

\begin{titlepage}
\PHyear{2024}       
\PHnumber{303}      
\PHdate{13 November}  

\title{Measurement of $\omega$ meson production in pp collisions at $\mathbf{\sqrt{\textit{s}} = 13\,\text{TeV}}$}
\ShortTitle{Measurement of $\omega$ meson production in pp collisions at $\sqrt{s}=\SI{13}{TeV}$}   

\Collaboration{ALICE Collaboration\thanks{See Appendix~\ref{app:collab} for the list of collaboration members}}
\ShortAuthor{ALICE Collaboration} 

\begin{abstract}
The \pt-differential cross section of $\omega$ meson production in pp collisions at $\sqrt{s}=\SI{13}{TeV}$ at midrapidity ($|y|<0.5$) was measured with the ALICE detector at the LHC, covering an unprecedented transverse-momentum range of $\num{1.6}<\pt<\SI{50}{GeV}/c$.
The meson is reconstructed via the $\omega\rightarrow\pi^+\pi^-\pi^0$ decay channel.
The results are compared with various theoretical calculations:
PYTHIA8.2 with the Monash 2013 tune overestimates the data by up to 50\%, whereas good agreement is observed with Next-to-Leading Order (NLO) calculations incorporating $\omega$ fragmentation using a broken SU(3) model.
The $\omega/\pi^0$ ratio is presented and compared with theoretical calculations and the available measurements at lower collision energies.
The presented data triples the \pt ranges of previously available measurements.
A constant ratio of $C^{\omega/\pi^0}=0.578\pm0.006~\text{(stat.)}\pm 0.013~\text{(syst.)}$ is found above a transverse momentum of $\SI{4}{GeV}/c$, which is in agreement with previous findings at lower collision energies within the systematic and statistical uncertainties.

\end{abstract}
\end{titlepage}

\setcounter{page}{2} 

\section{Introduction} 
Measuring single inclusive hadron production cross sections in high-energy hadronic collisions is an important tool to test our understanding of strong interactions and the underlying field theory of quantum chromodynamics (QCD)~\cite{QCD}.
For large enough momentum transfers $Q^2$, hadron production can be described in perturbative QCD (pQCD), where a ``hard'' point-like scattering produces a hadron via the fragmentation of an outgoing parton.
However, while the QCD matrix elements can be calculated within pQCD for sufficiently large scales, both the initial state of the collision as well as the fragmentation process are not amenable to a perturbative treatment, due to their complexity and the small momentum transfers involved.
Such calculations of high-\pt hadron production (as well as other high-$\pt$ observables) rely on the factorization~\cite{factorization} into three different contributions: 1. parton distribution functions (PDFs)~\cite{pdf} which describe the probability density to find a given parton with a given momentum inside the colliding hadrons, 2. QCD matrix elements describing the scattering of partons, and 3. fragmentation functions (FFs)~\cite{fragmentation} that relate the outgoing parton momentum to the momentum of the observed hadrons.
Both, FFs and PDFs cannot be calculated from first principles and rely on the determination from experimental data.
In addition, the majority of particles produced in a hadronic collision originate from soft scattering processes that involve small momentum transfers and therefore cannot be treated perturbatively and fully rely on phenomenological models that also require experimental verification.

Comparisons of measured hadron production cross sections with calculations are therefore essential to test and improve our understanding of the hadronization process and the initial state of the collision, i.e. to provide constraints for the FFs and PDFs, which are an indispensable ingredient for many experimental and theoretical analyses.
For example, recent measurements of $\pi^0$ and $\eta$ production~\cite{pi07TeV,pi0276,pi08TeV} provided new constraints for the nuclear gluon PDF~\cite{nCTEQ}, as well as the gluon-to-pion fragmentation function~\cite{FragFunctionALICEData}, where the gluon in particular drives a significant fraction of the scatterings relevant for hadron production.
Studies of $\omega$ meson production are especially interesting:
even though the $\omega$ meson consists mainly of light valence quarks like in the case of the $\pi^0$ and $\eta$, it has a larger mass of $\SI{782}{MeV}/c^2$ and carries spin $1$.
This makes $\omega$ meson production an interesting additional probe for parton fragmentation, where in particular comparisons with the data for other mesons can help to disentangle the role of spin, mass, and flavor in the hadronization process.
While a number of theoretical analyses focus on the fragmentation of pseudoscalar mesons and baryons such as $\pi$, K, $\eta$, and protons~\cite{fragmentationeffort1,fragmentationeffort2}, only a few models~\cite{omegafrag1,omegafrag2,omeganlo} describe fragmentation in the vector-meson sector -- mainly due to a lack of experimental data.
Here, experimental inputs for such studies are povided by presenting the most precise $\omega$ production results in hadronic collisions over an unprecedented transverse-momentum range measured with the ALICE experiment.

This article presents the invariant \pt-differential cross section of inclusive $\omega$ meson production at midrapidity ($|y|<0.5$) in pp collisions at $\sqrt{s}=\SI{13}{TeV}$. The measurements cover a momentum range of $\num{1.6}<\pt<\SI{50}{GeV}/c$, extending the previous ALICE measurement of $\omega$ production in pp collisions at $\sqrt{s}=\SI{7}{TeV}$~\cite{omega7TeV} by almost a factor of three towards higher \pt.
In addition, the $\omega/\pi^0$ ratio is presented,
which, e.g., allows to test the validity of $m_{\text{\tiny T}}$-scaling~\cite{mtscaling}, an empirical scaling rule widely used to obtain identified particle spectra in the absence of experimental data.
The $\omega$ production cross section as well as the $\omega/\pi^0$ ratio are confronted with theoretical predictions, using pQCD at NLO calculations based on the broken SU(3) model~\cite{omeganlo} and the PYTHIA8.2~\cite{pythia82} event generator, as well as with other experimental results at lower collision energies.

The paper is structured as follows: Section~\ref{sec:alicedetector}
briefly describes the detectors of the ALICE apparatus which are relevant for the presented measurements.
The event selection, as well as charged particle and photon reconstruction are outlined in Secs.~\ref{sec:eventandtrackselection} and~\ref{sec:photons}.
The $\omega$ meson reconstruction as well as acceptance and efficiency corrections are discussed in Sec.~\ref{sec:mesons}.
The evaluation of the systematic uncertainties, the presentation of the results, and comparison with theoretical models, as well as the conclusions are given in Secs.~\ref{sec:systematics},~\ref{sec:results} and~\ref{sec:conclusion}, respectively.

\section{ALICE detector}
\label{sec:alicedetector}
The $\omega$ meson is reconstructed via its decay into $\pi^+\pi^-\pi^0$ with a branching ratio $\text{BR}=(89.2\pm0.7)\%$, followed by $\pi^0\rightarrow\gamma\gamma$ with $(99.823\pm0.034)\%$~\cite{PDG}.
This necessitates the measurement of photons, as well as tracks from charged particles with the ALICE detector.
The charged pions are measured and identified using the ALICE central tracking system, which consists of the Inner Tracking System (ITS)~\cite{ITS} and the Time Projection Chamber (TPC)~\cite{TPC}.
Several methods exist to reconstruct photons, all of which are utilized in this work, using either the electromagnetic calorimeter (EMCal)~\cite{EMCALTDR,EMCalPerfPaper} or the photon spectrometer (PHOS)~\cite{phostdr}.
In addition, photon conversions into $\text{e}^+$$\text{e}^-$ pairs occurring within the central tracking system are exploited to reconstruct photons down to low transverse momenta.
All detectors relevant to this work are briefly outlined below, including the V0 detector~\cite{V0Detector}, which serves as a minimum bias trigger in this measurement.
A full description of the ALICE detector system and its performance can be found in Refs.~\cite{ALICE} and~\cite{ALICEPerformance}, respectively.

The ITS consists of two inner layers of Silicon Pixel Detectors (SPD), two layers of Silicon Drift Detectors (SDD), and two outermost layers of Silicon Strip Detectors (SSD).
The ITS is the detector located closest to the nominal interaction vertex, where the layers are positioned between \SI{3.9}{cm} and \SI{43.0}{cm} radial distance from the beam pipe.
The two SPD layers cover a pseudorapidity range of $|\eta|<2$ and $|\eta|<1.4$, respectively, whereas the SDD and SSD layers have a smaller coverage of $|\eta|<0.9$ and $|\eta|<1.0$, respectively.
The ITS is used for the tracking of charged particles and the reconstruction of the main collision vertex.

The TPC is a cylindrical drift detector, which consists of a large cylindrical field cage filled with $\SI{90}{m^3}$ of a gas mixture of Ar-CO$_2$ in proportions 90--10 in 2016 and 2018, and Ne-CO$_2$-N$_2$ (90--10--5) in 2017.
The active volume of the TPC ranges from about \SI{85}{cm} to \SI{250}{cm} in radial distance from the interaction point,  it covers a pseudorapidity window of $|\eta|<0.9$ over the full azimuthal angle.
It is the main tracking detector of ALICE and enables the measurement of charged-particle tracks with up to 159 space points each, as well as particle identification via the specific energy loss ($\dv*{E}{x}$) along their trajectory through the gas.
The large solenoid magnet surrounding the central barrel detectors provides a nominal magnetic field of $B=\SI{0.5}{T}$, which enables the reconstruction of charged tracks down to low transverse momenta ($\pt\approx\SI{100}{MeV}/c$).
By combining the track reconstruction in ITS and TPC, a transverse momentum resolution of about \mbox{1\% (3\%) is achieved at $\pt=\SI{1}{GeV}/c$ ($\SI{10}{GeV}/c$)~\cite{ALICEPerformance}.} 

The EMCal is a Pb-scintillator sampling calorimeter using wavelength-shifting fibers to collect the scintillation light from an alternating stack of 76 lead absorber and 77 scintillation layers.
The EMCal covers $|\eta|<0.7$ in pseudorapidity and $\Delta\varphi=\ang{107}$ in azimuthal angle, and consists of \num{12288} individual towers in total, each with a size of $\Delta\eta\cross\Delta\varphi\approx 0.0143\cross0.0143$, corresponding to roughly two times the Molière radius.
An extension of the EMCal located on the opposite side in azimuthal angle, which is referred to as the DCal, covers $0.22<|\eta|<0.7$ for $\ang{260}<\varphi<\ang{320}$ and  $|\eta|<0.7$ for $\ang{320}<\varphi<\ang{327}$, adding additional \num{5376} towers to the readout.
The EMCal enables the measurement of photons and electrons and the energy resolution is given by $\sigma_{E}/E=4.8\%/E\oplus 11.3\%/\sqrt{E}\oplus 1.7\%$  with energy $E$ in units of GeV~\cite{EMCalPerfPaper}.
In addition, the EMCal offers several hardware triggers which are used in this analysis and discussed in Sec.~\ref{sec:eventandtrackselection}.

The PHOS is an electromagnetic calorimeter based on lead-tungstate (Pb$\text{WO}_4$) scintillation crystals.
It has a smaller acceptance than the EMCal, covering $\Delta\varphi=\ang{70}$ and $|\eta|<0.12$. However, it offers a higher granularity, i.e. each of its 12544 crystals has a size of about $2.2\cross\SI{2.2}{cm}^2$, where the lateral dimension is only slightly larger than the Molière radius of Pb$\text{WO}_4$ of \SI{2}{cm}.
The PHOS is operated at a temperature of \SI{-25}{\celsius}, which ensures a high light yield of the crystals and consequently an energy resolution of $\sigma_E/E=1.3\%/E\oplus 3.6\%/\sqrt{E}\oplus 1.1\%$ with energy $E$ in units of GeV~\cite{PHOSCalibration}.

The V0 detector consists of two scintillator arrays located at forward and backward rapidity, covering pseudorapidties of $2.8<\eta<5.1$ and $-3.7<\eta<-1.7$, respectively.
It provides the minimum bias trigger and is used in this analysis to reduce background events originating from beam--gas interactions and out-of-bunch pileup.

\section{Event and track selection}
\label{sec:eventandtrackselection}
The $\omega$ measurement is performed using pp collision data at a center-of-mass energy of $\sqrt{s}=\SI{13}{TeV}$ recorded with the ALICE experiment from 2016 to 2018.
All considered collision events fulfill a minimum bias trigger condition, denoted as $\text{MB}$,  which requires the coincidence of signals in both V0 scintillator arrays.
The cross section of the minimum bias trigger was evaluated in van der Meer scans~\cite{vdm1,vdm2} and found to be $\sigma_\text{MB}=\SI{57.97\pm0.92}{mb}$~\cite{ALICELuminosity13TeV}.
In addition, two EMCal Level 1 (L1) hardware triggers~\cite{EMCalPerfPaper} are used, which provide a trigger decision after about \SI{6.5}{\micro s} and allow the selection of events with energy deposits above two configurable thresholds in the EMCal.
The energy deposits are evaluated in a sliding $4\cross 4$ tower window using dedicated Trigger Readout Units (TRUs)~\cite{EMCalPerfPaper}.
When the data used in this work was recorded, the EMCal L1 triggers operated at thresholds of \num{4} and \SI{9}{GeV} and are referred to as EG2 and EG1, respectively.
Both trigger classes used in this measurement are required to be exclusive and events fulfilling both the EG2 and EG1 trigger conditions are assigned to the lower threshold EG2 trigger.
The rejection factors of the EMCal triggers were evaluated using EMCal photon cluster-energy distributions, where the ratio of the spectra for the EMCal triggers with respect to the minimum bias trigger becomes constant above the trigger threshold and can be fitted to quantify the rejection power of each trigger.
For more details on this procedure, the reader is referred to Refs.~\cite{pi0eta8TeV} and~\cite{EMCalPerfPaper}.
The trigger rejection factors (RFs) and their corresponding systematic uncertainties were found to be \num{436.5\pm16.5} and \num{5443\pm208} for the EG2 and EG1 triggers, respectively.
Beam-induced background, such as beam--gas interactions and out-of-bunch pileup are rejected using the timing information provided by the V0 detectors, as well as the correlation between the number of hits and track segments in the SPD, where no correlation is expected for background events.
In-bunch pileup is rejected by requiring that only one primary collision vertex is reconstructed per event from track segments in the SPD.
In addition, events with a reconstructed vertex more than \SI{10}{cm} from the nominal collision point along the beam axis are rejected.
The nominal integrated luminosities $\mathcal{L}_{\text{int}}=N_{\text{evt}}\times\text{RF}/\sigma_{\text{MB}}$ of the measurement are $\mathcal{L}_{\text{int}}^{\text{MB}}=26-\SI{32}{nb}^{-1}$, $\mathcal{L}_{\text{int}}^{\text{EG2}}=958-\SI{985}{nb}^{-1}$, and $\mathcal{L}_{\text{int}}^{\text{EG1}}=8314-\SI{8550}{nb}^{-1}$ for the MB, EG2, and EG1 triggered samples, respectively.
The number of analyzed events ($N_{\text{evt}}$) is required to fulfill the previously outlined selections and is corrected for the vertex finding efficiency of 99.4\%.
The specified ranges indicate the dependence of the luminosity on the used $\pi^0$ reconstruction method (see Sec.~\ref{sec:mesons}), which arises due to slight differences in the used number of events ($N_{\text{evt}}$) depending on the involved detectors.

The trajectories (tracks) of charged particles with $|\eta|<0.9$ are reconstructed in the ITS and TPC, where at least 80 crossed pad rows in the TPC and at least one hit in any of the ITS layers were required for each track.
To ensure an overall good quality of the obtained tracks, the $\chi^2/$ndf of the tracks in the TPC was required to be below $4$ per track point and tracks with a transverse momentum below $\SI{100}{MeV}/c$ were rejected.
To identify tracks belonging to primary charged pions, the tracks were first loosely constrained to the collision vertex, requiring a distance of closest approach to the collision vertex within \SI{3.2}{cm} along the beam direction and \SI{2.4}{cm} in the transverse plane.
In addition, the specific energy loss $\dv*{E}{x}$ of the particle in the TPC is used to identify tracks from charged pions by requiring that its value lies within $3\sigma$ of the expected average $\dv*{E}{x}$ signal for charged pions.

\section{Photon measurement}
\label{sec:photons}
In contrast to the charged pions, the neutral pions produced in the  $\omega\rightarrow\pi^+\pi^-\pi^0$ channel, almost instantaneously decay with a branching ratio of $(99.823\pm0.034)\%$~\cite{PDG} to photon pairs, which necessitates the reconstruction of photons.
In this article, all available methods to reconstruct photons at midrapidity with the ALICE detector were exploited:
The EMCal\footnote{If not specified otherwise, references to the EMCal detector always include the DCal on the opposite side in azimuth angle.} and PHOS detectors are used to measure photons via the electromagnetic showers they produce in the detector elements, which allows to reconstruct photons above $E\sim\SI{0.5}{GeV}$.
In addition, the so-called photon conversion method (PCM) \cite{ALICEPerformance} is used, which allows to measure photons down to lower energies when they convert to $\text{e}^+\text{e}^-$ pairs within the inner detector material.

The electromagnetic shower produced by a particle interacting with the detector material of the EMCal or PHOS usually deposits its energy over multiple adjacent towers, which requires employing algorithms that combine these individual deposits into so-called clusters.
The algorithms are optimized to maximize reconstruction efficiency and shower separation and they are discussed in detail in Ref.\,\cite{EMCalPerfPaper}.
Clusters originating from minimum-ionizing particles as well as detector noise are suppressed by requiring for each reconstructed cluster in the EMCal (PHOS) a total energy of $E_{\text{clus}}>\SI{0.7}{GeV}$ (\SI{0.3}{GeV}) and to consist of at least  \num{2} towers for PHOS at energy deposits greater than \SI{1}{GeV}.
Since the time integration windows of the EMCal and the PHOS of about \SI{1.5}{\micro s} and \SI{3}{\micro s}, respectively, are larger than the LHC bunch spacing of \SI{25}{\nano s}, out-of-bunch pileup contributions are suppressed by requiring that the highest-energy tower of each cluster has an arrival time $t_\text{clus}$ of $-\SI{20}{\nano s }\leq t_{\text{clus}}\leq\SI{25}{\nano s}$  for the EMCal and $|t_{\text{clus}}|<\SI{30}{ns}$ for the PHOS measurement.
Clusters produced by photons are identified using the shape of the cluster as well as a track-matching veto.
The cluster shape can be parameterized in two dimensions using an ellipse, where the larger eigenvalue $\sigma^2_{\text{long}}$ of its dispersion matrix quantifies the elongation of the cluster.
For EMCal clusters, a requirement of $0.1\leq\sigma^2_{\text{long}}<0.7$ is imposed, where the lower threshold removes contaminations from, e.g.\,neutrons hitting the readout electronics, and the upper threshold suppresses elongated clusters originating from low-\pt electron and hadron tracks, as well as overlapping showers of multiple particles reconstructed as a single cluster~\cite{EMCalPerfPaper}.
Due to the higher granularity of the PHOS, only the lower threshold was found to be sufficient for the measurement and is applied for clusters with $E>\SI{1}{GeV}$.
The track-matching veto is performed using tracks reconstructed with the ITS and TPC, where clusters originating from charged particles can be suppressed by requiring that no track is pointing to the reconstructed cluster within a given $\Delta\phi$--$\Delta\eta$ window.

About 8.5\%~\cite{ALICEPerformance} of photons traversing the ALICE inner detector material convert to an $\text{e}^+\text{e}^-$ pair within a radial distance of \SI{180}{cm} from the nominal interaction point.
This allows to exploit ALICE's tracking capabilities to reconstruct the photon from the $\text{e}^+\text{e}^-$ tracks in the ITS and TPC within the fiducial acceptance of $|\eta|<0.9$ using the PCM.
The photon conversions are characterized by the distinct topology of two tracks with opposite curvature that orignate from a common point within the tracking detectors, often displaced by more than \SI{10}{cm} from the primary collision vertex.
These V shaped topologies of neutral two body decays ($\text{V}^0$s) are identified using a dedicated algorithm~\cite{ALICEPhysicsPerf}, which is also used in other measurements of, e.g., K$_{\rm S}^{0}$ and $\Lambda$ decays~\cite{lambda}.
Following some general tracking requirements to ensure an overall good reconstruction quality of the tracks originating from displaced secondary vertices, electron candidate tracks were identified via the specific energy loss $\dv*{E}{x}$ along their trajectory in the TPC.
In particular, electron candidate tracks were required to have a specific energy loss within $-3\sigma$ and $4\sigma$ of the expected energy loss of electrons, which is parametrized according to the Bethe-Bloch formula~\cite{ALICEPerformance,PID}.
In addition, contaminations from charged pion tracks were suppressed by rejecting tracks whose $\dv*{E}{x}$ were within $1\sigma$ of the expected energy loss for charged pions.
In order to ensure a good quality of the obtained $\text{V}^0$ sample, the reduced $\chi^2$ of the Kalman filter hypothesis~\cite{V0Kalman} for the photon candidate topology is constrained and the pair momentum vector is required to point towards the primary vertex.
Contributions from Dalitz decays are suppressed by rejecting $\text{V}^0$ candidates too close to the primary collision vertex.
Remaining contaminations, mainly from $K_S^0$, $\Lambda$, and $\overline{\Lambda}$ decays, are suppressed by studying the decay kinematics of the vertex candidate in an Armenteros--Podolanski diagram~\cite{podolanski}, where photon conversions can be identified as decays of a particle with vanishing rest mass.

\begin{figure}[h!]
    \centering
    \subfigure[PCM]{\includegraphics[width=0.37\textwidth]{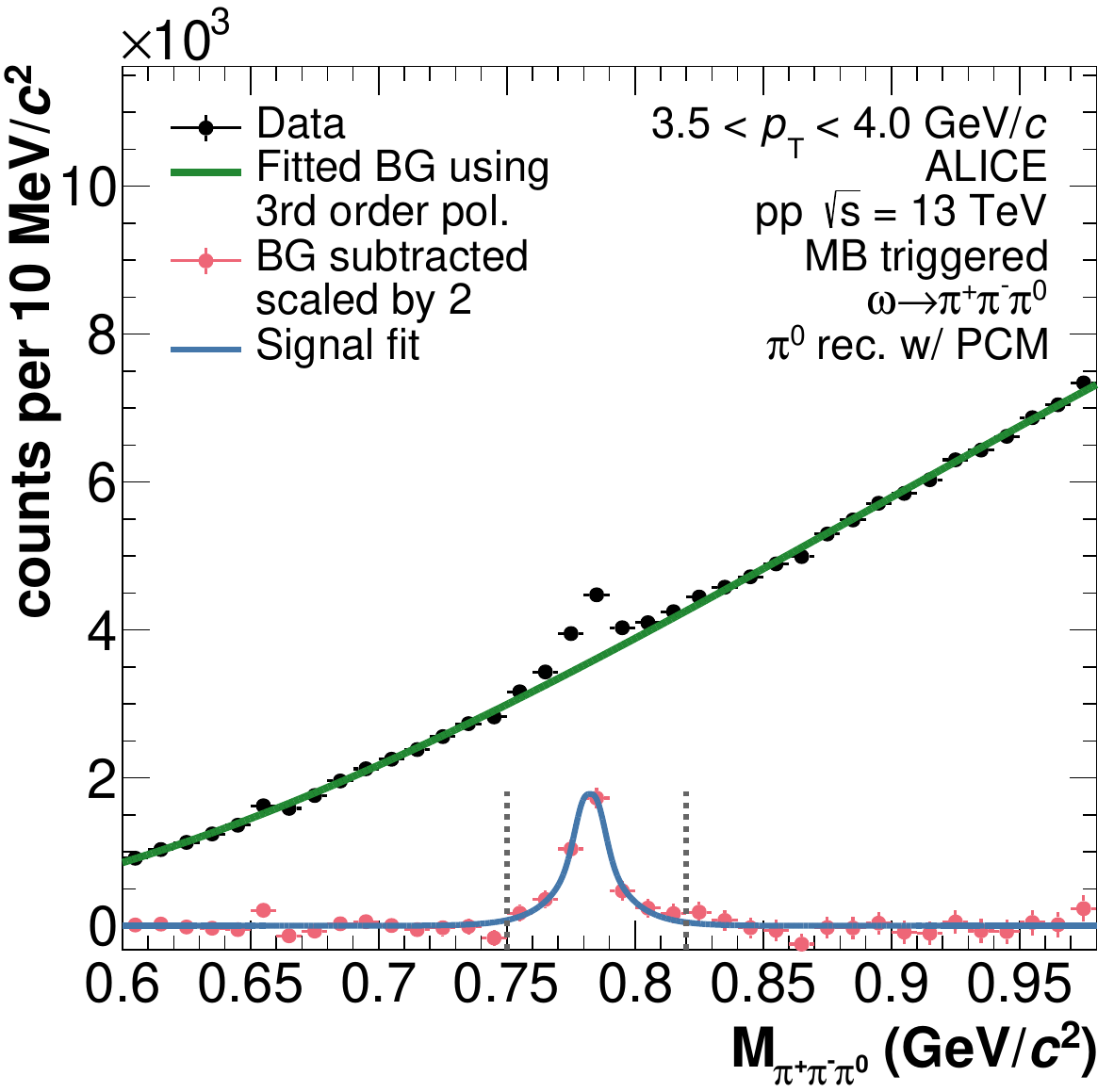}}
    \subfigure[PCM--EMC]{\includegraphics[width=0.37\textwidth]{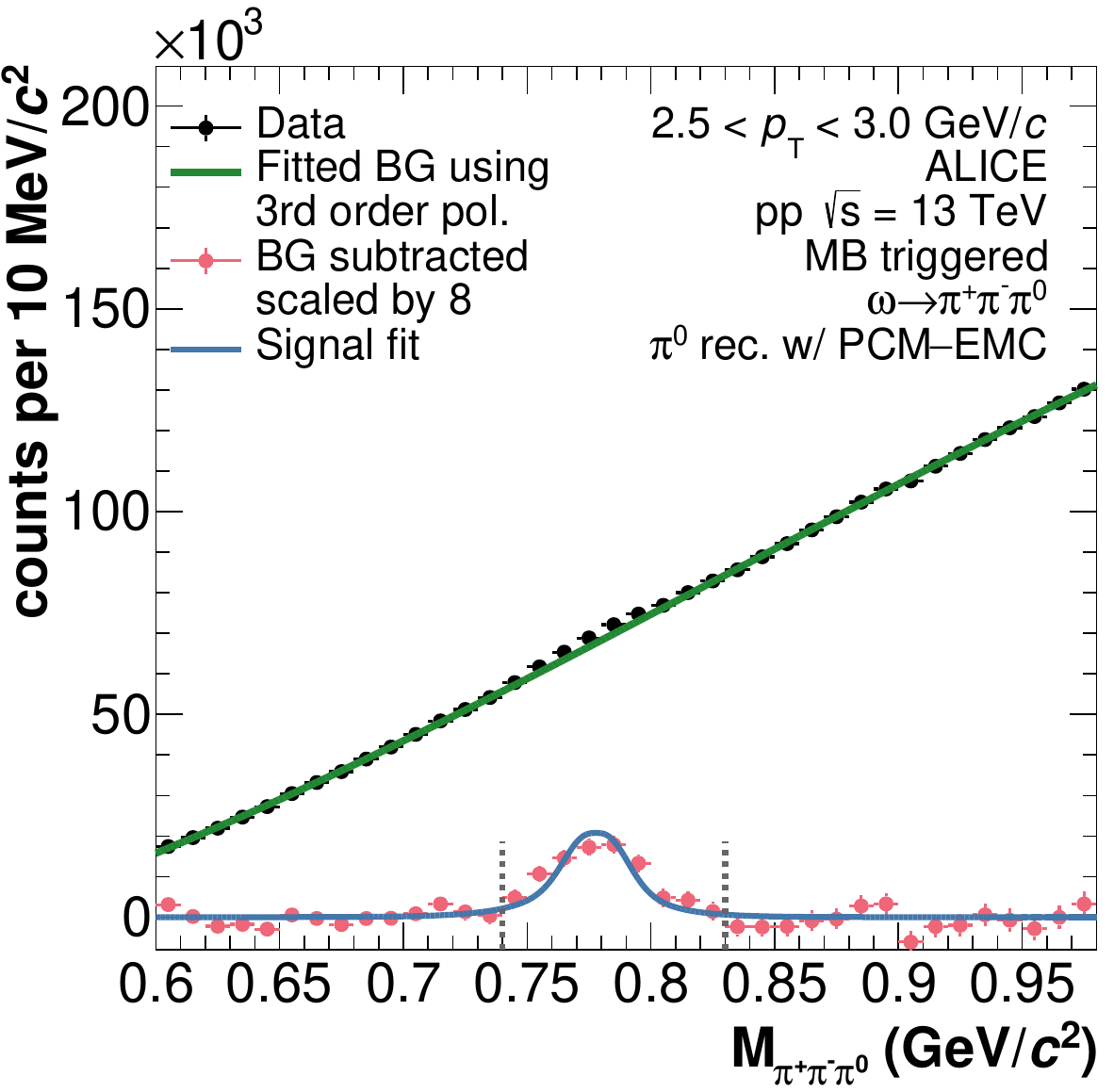}}
    \subfigure[PCM--PHOS]{\includegraphics[width=0.37\textwidth]{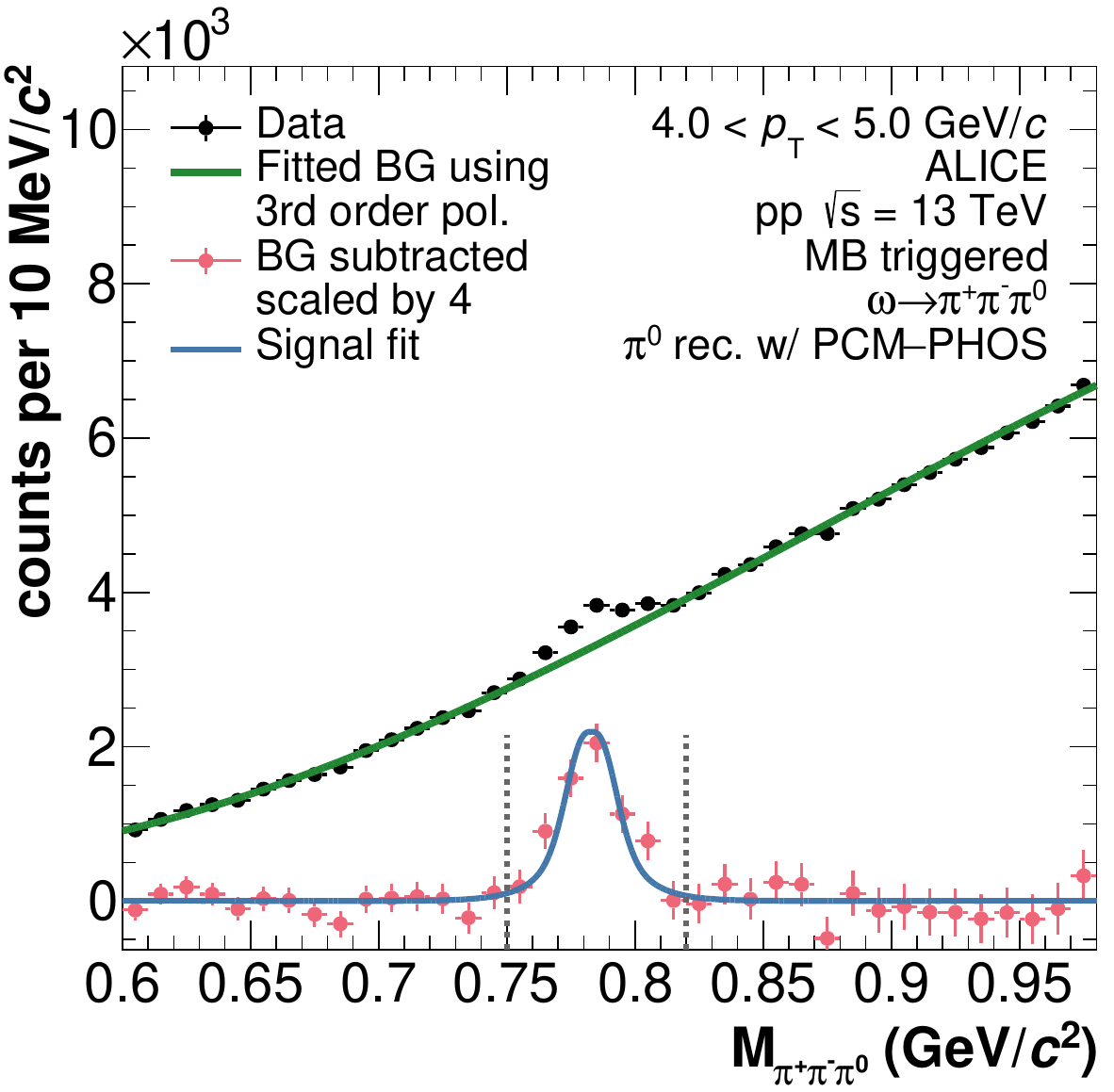}}
    \subfigure[PHOS]{\includegraphics[width=0.37\textwidth]{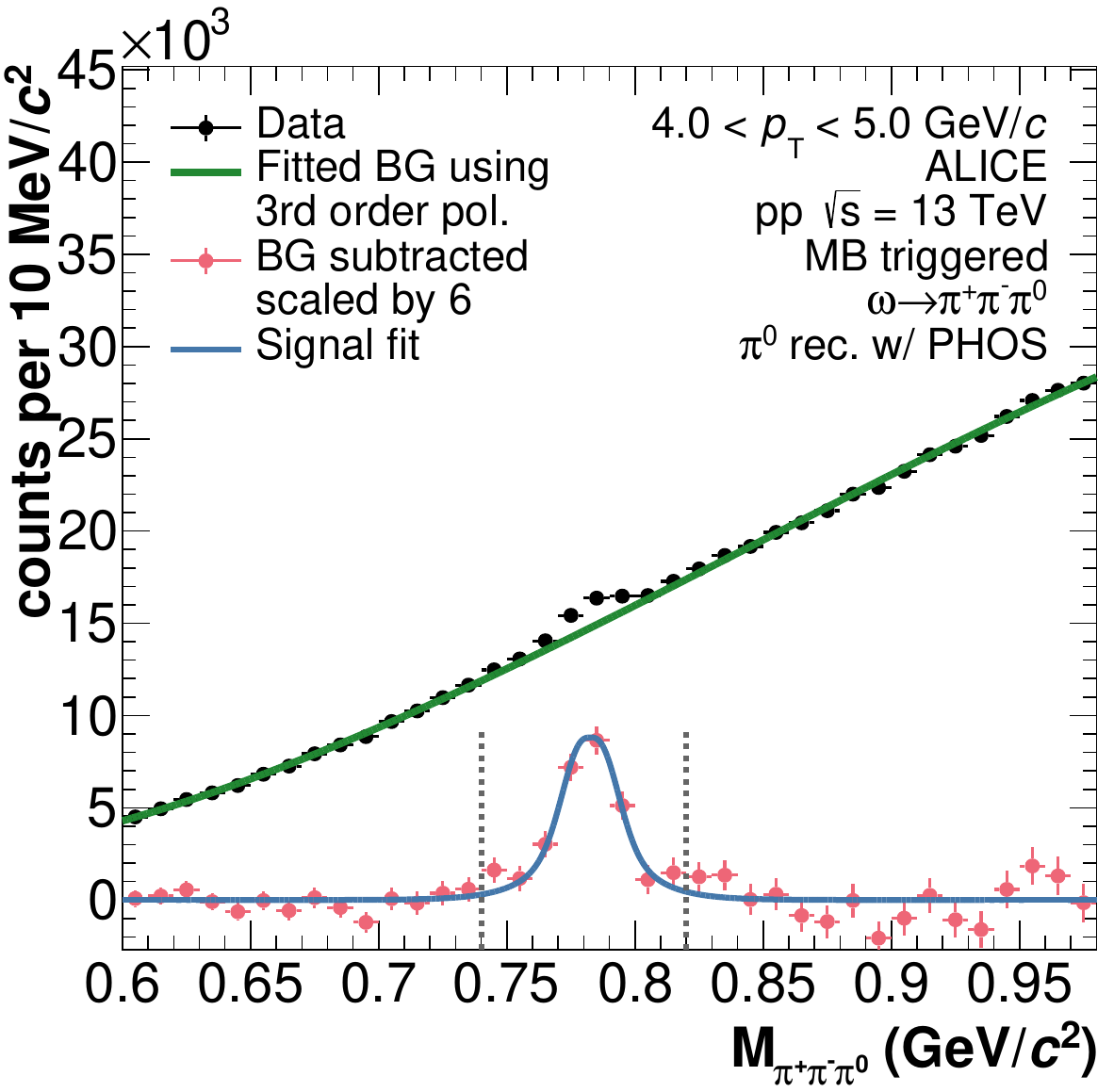}}
        \subfigure[EMC]{\includegraphics[width=0.37\textwidth]{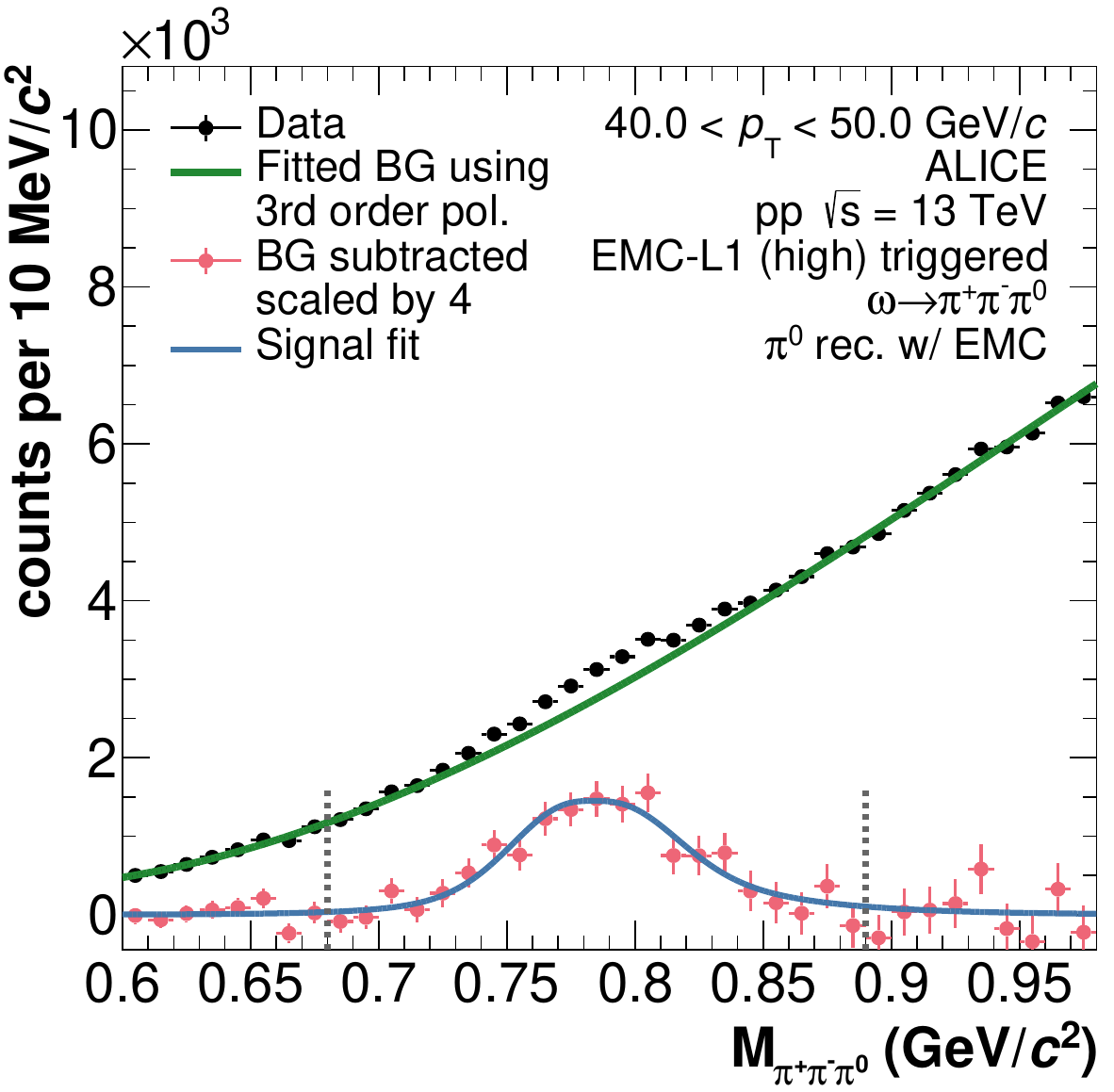}}
    \caption{Invariant mass distributions of $\pi^+\pi^-\pi^0$ candidates shown in the vicinity of the nominal $\omega$ mass in exemplary \pt intervals. Each panel displays a different $\pi^0$ reconstruction method, which is indicated in the respective panel and described in the text. A third-order polynomial is used to describe the background, which is subtracted from the distribution and the obtained signal is fitted with a Gaussian with two exponential tails. The integration range used to obtain the raw yields via bin-by-bin counting is indicated by vertical gray lines.}
    \label{fig:examplebins}
\end{figure}

\section{Meson reconstruction}
\label{sec:mesons}
Neutral pion candidates are reconstructed by calculating the two-photon invariant mass of all possible photon pairs ($M_{\gamma\gamma}$) in a given event that fulfill the selection criteria outlined in the previous section.
In this measurement, five different methods were used to reconstruct the $\pi^0$ candidates: 
for the methods referred to as PCM, PHOS and EMC in the following, \emph{both} photons from the $\pi^0$ decay were reconstructed using the PCM technique, the PHOS, and the EMCal detectors, respectively.
In addition, two hybrid approaches PCM--EMC and PCM--PHOS are used, where one photon was reconstructed using the PCM and the other with the respective calorimeter.
This allows to benefit from the high momentum resolution of the PCM as well as the high reconstruction efficiency of the calorimeters.
The obtained invariant mass distributions exhibit a peak in the number of photon pairs from $\pi^0$ decays, which is fitted for each method in $\pi^0$ \pt intervals using a Gaussian with mean $\mu_{\pi^0}$ and width $\sigma_{\pi^0}$.
$\pi^0$ candidates suitable for a subsequent $\omega$ reconstruction are then selected using an \pt-dependent invariant mass requirement of $\mu_{\pi^0}-3\sigma_{\pi^0}\leq M_{\gamma\gamma}\leq \mu_{\pi^0}+3\sigma_{\pi^0}$.

In order to reconstruct $\omega$ meson candidates, the invariant mass of all possible $\pi^+\pi^-\pi^0$ combinations is calculated in a given event, where the $\pi^0$ candidates are obtained according to the methods outlined above.
The nominal $\pi^0$ mass $m_{\pi^0}\approx \SI{134.98}{MeV}/c^2$~\cite{PDG} is assigned to all neutral pion candidates, which was found to improve the overall $\omega$ mass resolution by about 30\%, which subsequently improves the significance of the signal.
These improvements are taken into account in the statistical subtraction of the background and efficiency calculation.
Figure~\ref{fig:examplebins} shows the resulting invariant mass distributions for all five methods in exemplary \pt intervals, where clear peaks from $\omega$ decays are visible on top of a combinatorial background.
The background is described using a third-order polynomial, which is subtracted from the measurement.
The obtained signal is fitted using a Gaussian with two exponential tails.
The reconstructed omega mass ($\mu_\omega$) and width $(\sigma_{\omega})$ are identified with the mean and width of the fit, respectively.
The $\omega$ mass resolution is $\text{FWHM}\approx\SI{30}{MeV}/c^2$ and the reconstructed $\omega$ mass is found to be in agreement with the nominal mass of $m_{\omega}^{\text{PDG}}=\SI{782.66\pm0.13}{MeV}/c^2$~\cite{PDG}, where a slight dependence on \pt and reconstruction technique is observed.
The signal width contains the intrinsic decay width of $\Gamma_{\omega}\approx\SI{8}{MeV}$~\cite{PDG}, as well as detector resolution effects from the photon reconstruction and charged particle tracking.
Finally, the raw $\omega$ yields are obtained by the sum of counts of the background subtracted distribution within $\mu_{\omega}\pm3\sigma_\omega$ for each \pt interval and reconstruction method.

The geometrical acceptance as well as the reconstruction efficiency for each method are evaluated using Monte Carlo simulations of full pp collision events using the PYTHIA8.2~\cite{pythia82} event generator.
The $\omega$ three-body decay kinematics is adequately described by PYTHIA, which takes into account the experimentally observed phase space density distributions~\cite{omegadecay1,omegadecay2} in form of a weighting.
The produced particles are propagated through the ALICE detector using GEANT3 \cite{GEANT3}, which simulates material interactions taking into account the geometry and operation conditions of the detectors at the time of data taking.
Furthermore, the energy scale and the non-linearity of the PHOS and EMCal calorimeters are taken into account by tuning the Monte Carlo to ensure agreement of the reconstructed $\pi^0$ mass and width with data for each method.
The trigger efficiencies of the EG1 and EG2 triggers are included in the reconstruction efficiency calculation and range from about 97\% to 99\% \cite{EMCalPerfPaper}.
Good agreement between data and Monte Carlo is observed for all relevant cluster and track properties, as well as the reconstructed $\omega$ mass and width, within the statistical uncertainties of the measurement.
\begin{figure}[t]
    \centering
    \includegraphics[width=0.52\textwidth]{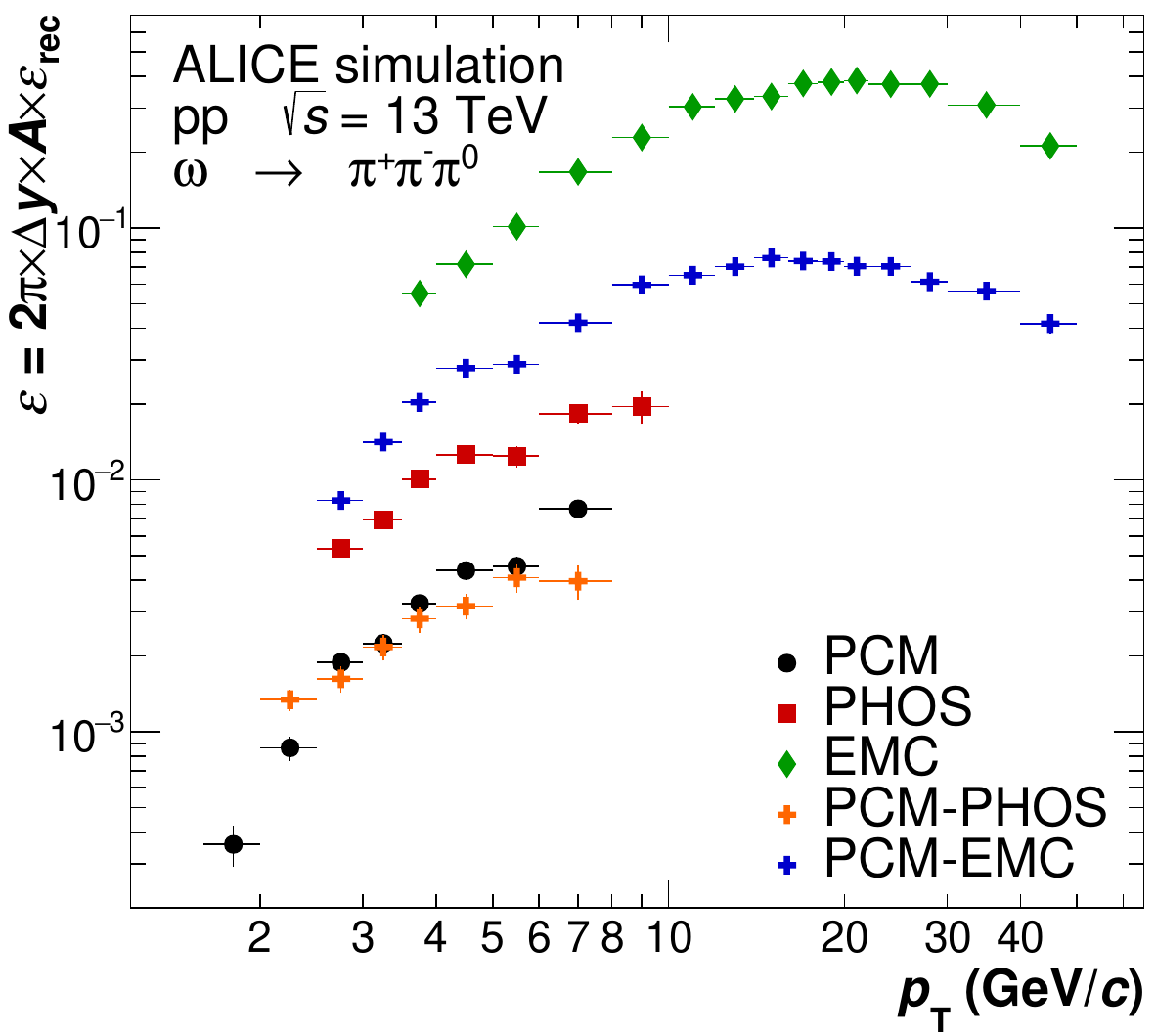}
    \caption{Correction factors applied to the raw $\omega$ yields according to Eq.~\ref{eq:xsection} for the five $\pi^0$ reconstruction methods, as indicated in the legend. The factors include the reconstruction efficiency $\epsilon_{\text{rec}}$ and the geometrical acceptance $A$ of the involved detectors. 
    }
    \label{fig:acceptanceefficiency}
\end{figure}
The full correction factors, $\epsilon$, which are applied for each reconstruction method and contain the efficiency as well as the geometrical acceptance are shown in Fig.~\ref{fig:acceptanceefficiency}.
In addition, a normalization to the rapidity interval, $\Delta y$, and $2\pi$ azimuthal angle is applied to allow for a direct comparison between the different methods.
The correction factors showcase the strength and \pt reach of each method, motivating the use of all available photon reconstruction methods at midrapidity in \mbox{ALICE}:
the PCM offers the lowest \pt reach of all methods, however, its efficiency is limited due to the low conversion probability of 8.5\%.
The two calorimeters extend the measurements to higher \pt, where in particular the use of the two EMCal L1 triggers allows reaching unprecedented transverse momenta up to $\SI{50}{GeV}/c$.
For each reconstruction method, the different triggers are appropriately combined on the level of fully corrected cross sections using the BLUE~\cite{BLUE} algorithm, taking into account the respective statistical and systematic uncertainties through a \pt-dependent weighting procedure.

\section{Systematic uncertainties}
\label{sec:systematics}
An overview of the systematic and statistical uncertainties of the measurement is given in Table~\ref{tab:syst}, where the uncertainties are evaluated on the $\omega$ production cross sections for the five different $\pi^0$ reconstruction methods and three trigger samples.
\begin{sidewaystable}
	\centering
		\caption{Overview of the relative systematic uncertainties in percent entering the measurement, which are given for each $\pi^0$ reconstruction method and trigger sample. Most uncertainties were found to be \pt dependent, and the uncertainties are quoted as ranges from lowest to highest uncertainty in the \pt-range inspected by the given method. The uncertainties are obtained as described in the text. The individual measurements and trigger samples are combined using the BLUE method~\cite{BLUE}, where each measurement is weighted according to its statistical and systematic uncertainties. The statistical and systematic uncertainties of the combined measurement are given in the bottom two rows.}
	\begin{tabular}{c|| c|c|c|c|c|c|c|c|c} 
		\toprule
		Method &  &  & PCM-- & \multicolumn{3}{c|}{} & \multicolumn{3}{c}{PCM--} \\
		 & PCM & PHOS & PHOS & \multicolumn{3}{c|}{EMC} & \multicolumn{3}{c}{EMC} \\
		\midrule
		Trigger & MB & MB & MB & MB & EG2 & EG1 & MB & EG2 & EG1\\
		\midrule
        Signal Extraction        & $3.7$--$15.5$    & $9.5$--$27.9$    & $4.5$--$27.1$    & $4.0$--$12.8$    & $1.9$--$8.3$     & $4.2$--$16.0$    & $5.2$--$16.4$    & $3.8$--$11.1$    & $5.4$--$22.3$  \\
        Material  budget               & $5.0$             & $2.0$             & $2.7$             & $4.2$             & $4.2$             & $4.2$             & $3.3$             & $3.3$             & $3.3$  \\
        Charged pion rec.        & $3.5$--$4.4$     & $3.7$--$4.4$     & $3.9$--$4.3$     & $4.2$--$4.6$     & $4.6$--$5.6$     & $4.8$--$6.0$     & $3.9$--$4.4$     & $4.3$--$5.4$     & $4.5$--$5.8$  \\
        Conv. photon rec.         & $1.9$--$9.9$     & -                 & $2.5$--$4.6$     & -                 & -                 & -                 & $2.3$             & $2.3$--$2.5$     & $2.3$--$2.8$  \\
        Calo photon rec.         & -                 & $6.1$--$6.4$     & $5.4$--$5.5$     & $2.5$--$3.5$     & $2.6$--$4.2$     & $4.8$--$6.9$     & $2.2$--$7.2$     & $2.2$--$5.1$     & $4.7$--$8.2$  \\	
        Neutral pion rec.        & $3.2$             & $5.0$--$5.6$     & $5.0$--$5.3$     & $3.1$             & $3.1$--$6.0$     & $3.2$--$10.0$    & $3.1$             & $3.1$--$5.8$     & $3.2$--$9.1$  \\
        Pileup                   & $0.5$             & $0.7$             & $0.7$             & $0.3$             & $0.3$             & $0.3$             & $0.3$             & $0.3$             & $0.3$  \\
        \midrule
        Total syst. uncertainty  & $8.4$--$21.7$    & $13.2$--$29.4$   & $10.8$--$29.0$   & $8.5$--$15.0$    & $7.7$--$12.7$    & $9.6$--$21.4$    & $9.1$--$19.3$    & $8.2$--$15.0$    & $10.1$--$23.8$  \\
        \midrule
        \midrule
        Statistical uncertainty  & $9.4$--$35.2$    & $9.3$--$19.5$    & $15.0$--$22.3$   & $4.5$--$7.9$     & $3.7$--$14.9$    & $3.9$--$9.6$     & $5.5$--$19.5$    & $7.3$--$24.4$    & $9.9$--$18.9$  \\
        \midrule
        \midrule
        Combined stat. unc.       & \multicolumn{9}{c}{$3.1$--$35.2$} \\
        Combined syst. unc.       & \multicolumn{9}{c}{$5.8$--$21.7$}  \\
        \bottomrule
	\end{tabular}
	\label{tab:syst}
\end{sidewaystable}	

The overall dominant contribution to the systematic uncertainty of each reconstruction method is the signal extraction uncertainty. 
It contains the uncertainty from the yield extraction, where the integration range was varied from $2\sigma$ to $4\sigma$, as well as several variations related to the description of the signal and background shape.
The latter includes variations of the background fit function and of the fitting range.
For the background description, the third-order polynomial is varied to a fourth-order polynomial.
In addition, the nominal Gaussian used to account for the peak region when subtracting the background is varied to include exponential tails on both sides. 
To account for a possible mismatch between the amount of material present in the ALICE detector and its implementation in GEANT3, a so-called material budget uncertainty is assigned for each measurement, which depends on the photon reconstruction method used to reconstruct the neutral pion candidates.
For measurements involving the PCM method, this uncertainty was found to be 2.5\% for single photons~\cite{materialbudget}.
This is an improvement from the previously assigned 4.5\%~\cite{pi07TeV}, which is achieved using a novel material weighting procedure presented in Ref.\,\cite{materialbudget}.
For EMCal photons, an uncertainty of 2.1\% per photon is assigned~\cite{pi0276}, which arises due to conversions in the detector material of the Time-Of-Flight (TOF) \cite{TOF} and Transition Radiation Detector (TRD) \cite{TRD}, which are positioned directly in front of the calorimeter.
The PHOS is only partially covered by TRD and TOF modules, resulting in a lower material budget uncertainty of 1\% for single photons, which was estimated using reconstructed $\pi^0$ candidates in different magnetic field configurations.
The resulting material budget uncertainty for both photons entering the $\pi^0$ reconstruction is given in Table~\ref{tab:syst}, where full correlation (PCM, EMC, and PHOS) or partial correlation (PCM-EMC, PCM-PHOS) has been assumed between the material budget uncertainties of the respective photons.
Finally, an uncertainty of 0.8\% is assigned to account for the uncertainties of the $\pi^0\rightarrow\gamma\gamma$ and $\omega\rightarrow\pi^+\pi^-\pi^0$ branching ratios.
It is obtained as the quadratic sum of the uncertainties of the $\pi^0$ and $\omega$ branching ratios using the PDG values
$(89.2\pm0.7)\%$ and $(98.823\pm0.034)\%$, respectively~\cite{PDG}.

The charged pion reconstruction uncertainty was evaluated through variations of the selection criteria outlined in Sec.\,\ref{sec:eventandtrackselection}, where on average the biggest contribution was found to be the uncertainty of the specific energy loss $\dv*{E}{x}$ selection to identify charged pion candidates ($\approx 2.5\%$), as well as the uncertainty of the ITS--TPC matching efficiency.
The latter was determined in Ref.\,\cite{ALICE:2020nkc} and increases with increasing track $\pt$ up to $\approx 5\%$ for the highest covered momenta.

The systematic uncertainty arising due to the photon reconstruction was estimated through independent variations of the selection criteria discussed in Sec.\,\ref{sec:photons}.
Using the PCM for photon reconstruction, the biggest uncertainty source was estimated to be on average on the order of 2\%, arising due to the selections applied to suppress secondary contaminations from K$_{\rm S}^{0}$, $\Lambda$, and $\overline{\Lambda}$ decays.
For the measurements involving the EMCal, the photon reconstruction systematic uncertainty is dominated  at low \pt by the minimum cluster energy requirement, whereas for larger \pt the uncertainty arising due to the determination of the EMCal trigger rejection factors becomes dominant.
The latter was estimated using the uncertainty of the rejection factor fit extracted from the cluster spectrum ratios.
For the photon reconstruction using the PHOS, the minimum cluster energy requirement and the selection according to the cluster time dominate the photon reconstruction uncertainty, each contributing with a systematic uncertainty of about 3\%.

The systematic uncertainty of the neutral pion reconstruction is dominated by the invariant mass requirement used to select $\pi^0$ candidates from all photon pairs in a given event.
The width of the selection window was varied between $2\sigma$ and $4\sigma$ around the reconstructed $\pi^0$ mass, resulting in an uncertainty on the order of $5\%$ on the fully corrected cross section.

Finally, the impact of in-bunch pileup on the measurement was evaluated by loosening the pileup-related selection criteria discussed in Sec.~\ref{sec:eventandtrackselection}, resulting in a \pt-independent 0.7\% systematic uncertainty.

In addition to the systematic uncertainties of each individual source, the total statistical and systematic uncertainties of each measurement are also given in Table~\ref{tab:syst}, where the latter is obtained as the quadratic sum of the uncertainties from the individual sources.
Correlations between individual measurements were taken into account and estimated to be about 30\% of the respective total systematic uncertainties.
The normalisation uncertainty of 1.58\% is fully correlated as a function of \pt and is given by the uncertainty of the minimum bias trigger cross section, which has been previously determined in Ref.~\cite{ALICELuminosity13TeV}.

\begin{figure}[t]
    \centering
    \includegraphics[width=0.8\textwidth]{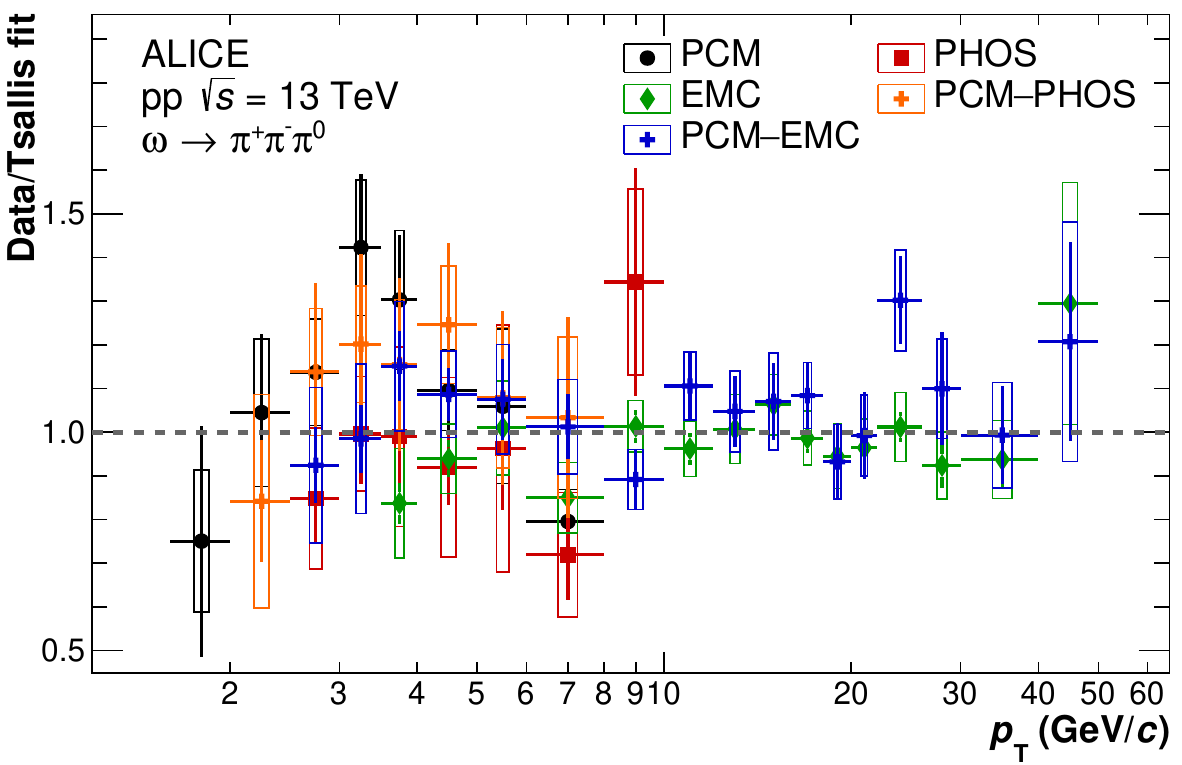}
    \caption{Ratios of the $\omega$ production cross section obtained using five different $\pi^0$ reconstruction methods with respect to a Levy--Tsallis fit of the combined measurement. The fit parameters are given in Table~\ref{tab:fitparameters} together with the reduced $\chi^2$ of the converged fit. The statistical and systematic uncertainties of the individual measurements are given in Table~\ref{tab:syst} and are represented by vertical bars and boxes, respectively.}
    \label{fig:combination}
\end{figure}
\section{Results}
\label{sec:results}
The invariant cross section of $\omega$ production  is calculated for the five reconstruction methods as
\begin{equation}
    E\dv[3]{\sigma^{\text{p}+\text{p}\rightarrow\omega+X}}{p}=\frac{1}{2\pi}\frac{1}{\pt}\times\frac{1}{\mathcal{L}_{\text{int}}}\times\frac{1}{A\times\epsilon_{\text{rec}}}\times\frac{1}{\kappa_{\text{BR}}}\frac{N^\omega}{\Delta y\Delta\pt},
    \label{eq:xsection}
\end{equation}
where $\mathcal{L}_{\text{int}}$ is the integrated luminosity given in Sec.~\ref{sec:eventandtrackselection}, $A$ and $\epsilon_{\text{rec}}$ are the geometrical acceptance and the reconstruction efficiency, respectively, which are determined for each individual reconstruction method and are shown in Fig.~\ref{fig:acceptanceefficiency}.
Furthermore, $\kappa_{\text{BR}}$ is the product of the branching ratios of the $\omega\rightarrow\pi^+\pi^-\pi^0$ and $\pi^0\rightarrow\gamma\gamma$ decays, which are
$(89.2\pm0.7)\%$ and $(98.823\pm0.034)\%$, respectively~\cite{PDG}.
The number of reconstructed $\omega$ meson candidates is denoted by $N^{\omega}$ and normalized to the transverse momentum interval width $\Delta\pt$, $2\pi$ azimuthal angle and rapidity interval width $\Delta y$.
The transverse momentum of the omega meson (\pt) in a given interval is corrected according to the underlying spectrum, as described in more detail below.

\begin{table}[b]
	\centering
	\newcommand{\UnitLumi}{nb$^{-1}$}
	\caption{Parameters and $\chi^2/\text{ndf}$ of the Levy--Tsallis~\cite{Tsallis} parametrization, which is given in Eq.~\ref{eq:tsallis} and was fitted to the invariant $\omega$ production cross section, as shown in Fig.~\ref{fig:crosssection}. The quoted statistical uncertainties are obtained by excluding the systematic uncertainties of the data points in the fitting procedure.}
	\begin{tabular}{ccccc} 
		\toprule
		$C$($\cross10^{10}$pb) & $n$ & $T$(GeV) & $\chi^2$/ndf & ndf \\
		\midrule
		$3.51^{\pm\text{0.63 (stat.)}}_{\pm\text{1.04 (tot.)}}$ & $6.17^{\pm0.05\text{ (stat.)}}_{\pm0.10\text{ (tot.)}}$ & $0.194^{\pm0.010\text{ (stat.)}}_{\pm0.020\text{ (tot.)}}$ & $^{1.92\text{ (stat.)}}_{0.46\text{ (tot.)}}$ & $16$  \\
		\bottomrule
	\end{tabular}
	\label{tab:fitparameters}
\end{table}	
The production cross section is measured for each reconstruction method and combined using the Best Linear Unbiased Estimate (BLUE) algorithm~\cite{BLUE}, which assigns \pt-dependent weights for each method, taking into account their statistical and systematic uncertainties.
Correlations of the systematic uncertainties between methods are taken into account and mainly originate from the material budget, signal extraction, and charged pion selection uncertainty, the latter of which is shared by all methods.
The statistical and systematic uncertainty of the combined measurement is given in the bottom rows of \mbox{Table~\ref{tab:syst}}, where uncertainties as low as 3.1\% and 5.8\%, respectively, are estimated.
\begin{figure}[t]
    \centering
    \includegraphics[width=0.6\textwidth]{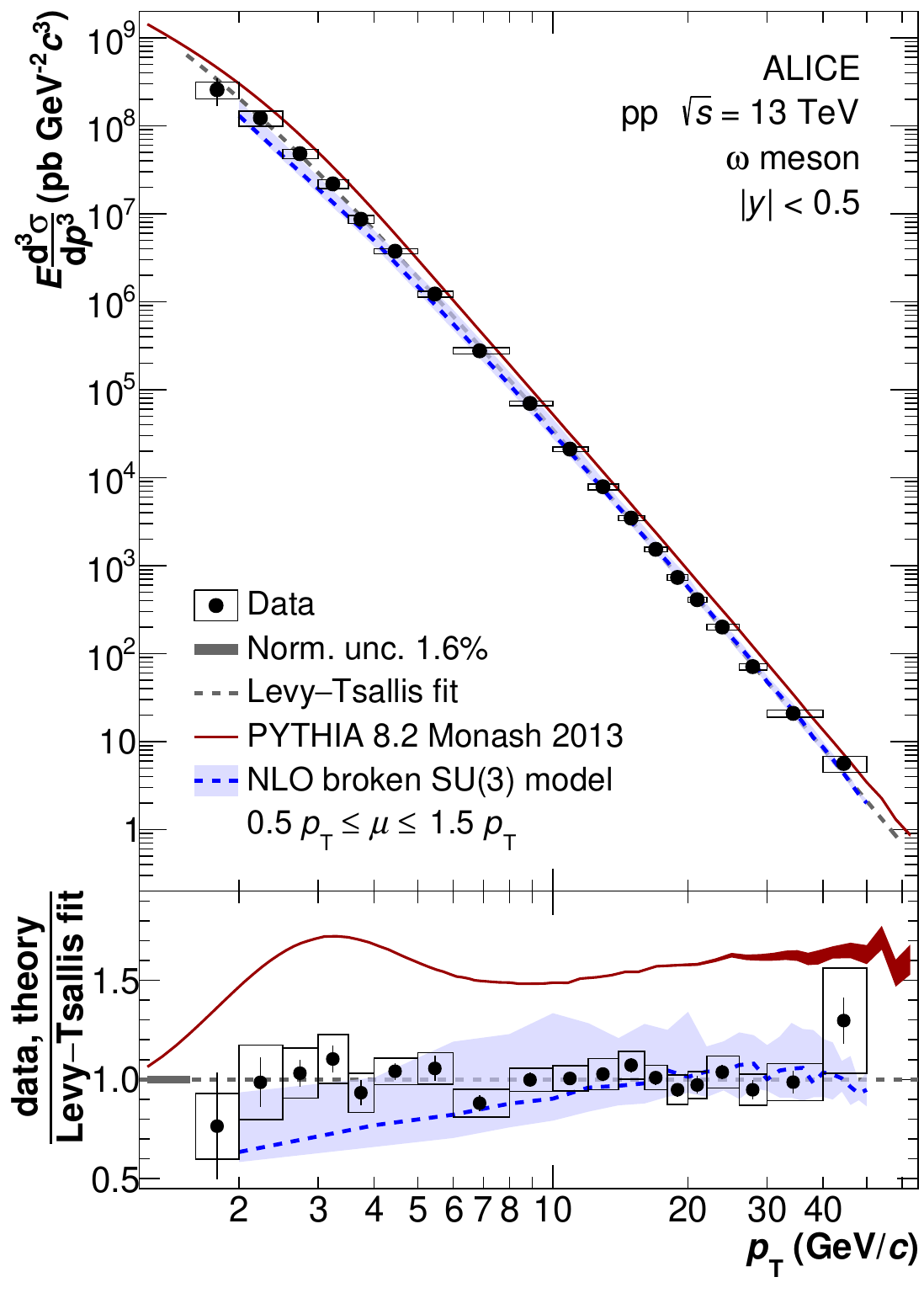}
    \caption{Invariant cross section of $p+p\rightarrow\omega+X$ production at midrapidity in pp collisions at $\sqrt{s}=\SI{13}{TeV}$ compared to theoretical predictions. The cross section is parametrized using a Levy--Tsallis function (dashed grey line), where the fit parameters are given in Table~\ref{tab:fitparameters}. 
    The systematic uncertainty of the measurement is denoted by boxes, excluding the normalization uncertainty of 1.58\%, which is shown separately as a grey box in the bottom panel of the figure.
    To account for the finite width of each \pt-interval, the data points are shifted according to the underlying spectrum, as described by the fit.
    The red line represents the theoretical prediction obtained using the PYTHIA8.2~\cite{pythia82} event generator with the Monash 2013 tune~\cite{Monash2013}. The width of the line indicates the statistical uncertainty of the prediction. The blue dashed line shows a NLO calculation~\cite{omeganlo} incorporating $\omega$ fragmentation based on a broken SU(3) model, where all scales are chosen to be $\mu=\pt$ and the blue shaded band denotes the scale uncertainty.
    The bottom panel shows the ratio of the data, as well as the theoretical predictions to the Levy--Tsallis parametrization.}
    \label{fig:crosssection}
\end{figure}
This is a reduction by a factor of two with respect to a previous ALICE publication on $\omega$ meson production in pp collisions at $\sqrt{s}=\SI{7}{TeV}$~\cite{omega7TeV}, which is mainly achieved by the inclusion of additional reconstruction methods and the reduced material budget uncertainty~\cite{materialbudget}.
Agreement between individual reconstruction techniques is illustrated in Fig.~\ref{fig:combination}, which shows the ratio of the individual cross sections with respect to a fit of the combined measurement.
In particular, a Levy--Tsallis function~\cite{Tsallis} is used to describe the combined spectrum, which is given by
\begin{equation}
    E\dv[3]{\sigma}{p}= \frac{C}{2\pi}\frac{(n-1)(n-2)}{nT\qty[nT+m(n-2)]}\qty(1+\frac{m_{\text{\tiny T}}-m}{nT})^{-n},
\label{eq:tsallis}
\end{equation}
where $m$ and $m_{\text{\tiny T}}=\sqrt{m^2+p_{\text{\tiny T}}^2}$ refer to the $\omega$ mass and transverse mass, respectively, and $C$,$T$ and $n$ are free parameters of the function.
The function is found to describe $\omega$ production over the full covered momentum range and its parameters are given in Table~\ref{tab:fitparameters} together with the reduced $\chi^2$ of the fit.
The individual reconstruction techniques agree with the combined fit within $1.9\sigma$ of the respective total experimental uncertainties.

The invariant production cross section of $\text{p}+\text{p}\rightarrow\omega+X$ production at midrapidity in pp collisions at $\sqrt{s}=\SI{13}{TeV}$ is shown in Fig.~\ref{fig:crosssection}.
The measurement covers an unprecedented transverse momentum range of $1.6<\pt<\SI{50}{GeV}/c$, extending previous measurements by over $\SI{30}{GeV}/c$.
To account for the finite width of the \pt-intervals, the shown cross section points are shifted from the bin center according to the method described in Ref.\,\cite{binshift}.
The $\omega$ meson spectrum is parametrized by the Levy--Tsallis function and the resulting shifts in \pt are found to be at most 2\%.
The data is confronted with two theoretical predictions and their ratio to the Levy--Tsallis fit of the measurement is shown in the bottom panel of the figure.
The PYTHIA8.2~\cite{pythia82} event generator using the Monash 2013 tune~\cite{Monash2013}  is shown as a red solid line, where the width of the line indicates the statistical uncertainty of the generated sample.
In the Monash 2013 tune, NNPDF2.3 LO~\cite{NNPDF23} is used to describe the incoming proton beams and hadrons are produced in a parton shower using the Lund string fragmentation model~\cite{lund}.
PYTHIA8.2 overestimates the measured cross section by about 50\% over almost the full covered momentum range.
This is in line with measurements of $\pi^0$ production~\cite{WAITINGPI0} and $\pi^\pm$ production~\cite{chargedpions} in the same collision system, where PYTHIA8.2  was found to overestimate the data by a similar amount at high \pt. 
Previous findings of $\omega$ meson production at lower collision energies of $\sqrt{s}=\SI{7}{TeV}$~\cite{omega7TeV} are consistent with PYTHIA8.2 predictions using the Monash 2013 tune, whereas a similar overestimation is observed for the older Tune 4C~\cite{Tune4C}.
This might indicate that the Monash 2013 tune, which was published in 2014 using early LHC data, is no longer sufficient to describe light meson production at the unprecedented collision energies, which have become accessible in the recent years of LHC operation.
In addition, the measurement is compared to next-to-leading order (NLO) calculations~\cite{omeganlo}, where $\omega$ fragmentation functions in vacuum are obtained  by evolving NLO DGLAP evolution equations~\cite{Q2evolution} with rescaled $\omega$ fragmentation functions from a broken SU(3) model.
This theoretical model allows to describe the fragmentation of vector mesons by reducing the parton fragmentation functions to only three independent functions by utilising SU(3) flavor symmetry, as well as isospin and charge conjugation invariance \cite{omegafrag1,omeganlo}.
Several free parameters of the model are fitted to experimental data from both RHIC and the LHC~\cite{omeganlo}.
The factorization scale $\mu$, renormalization scale $\mu_{\rm R}$ and fragmentation scale $\mu_{\rm F}$ are chosen to coincide with the $\pt$ of the $\omega$ meson.
The scale uncertainty was evaluated by varying the scales simultaneously from $0.5\pt$ to $1.5\pt$, as indicated by the blue shaded band in Fig.~\ref{fig:crosssection}.
The proton beams are described using the CT14~\cite{CT14} parton distribution functions.
The NLO calculation describes the measurement within the uncertainties over the whole covered transverse momentum range, where especially a good agreement is observed for $\pt^\omega\gtrsim\SI{8}{GeV}/c$.

\begin{figure}[t]
    \centering
    \includegraphics[width=\textwidth]{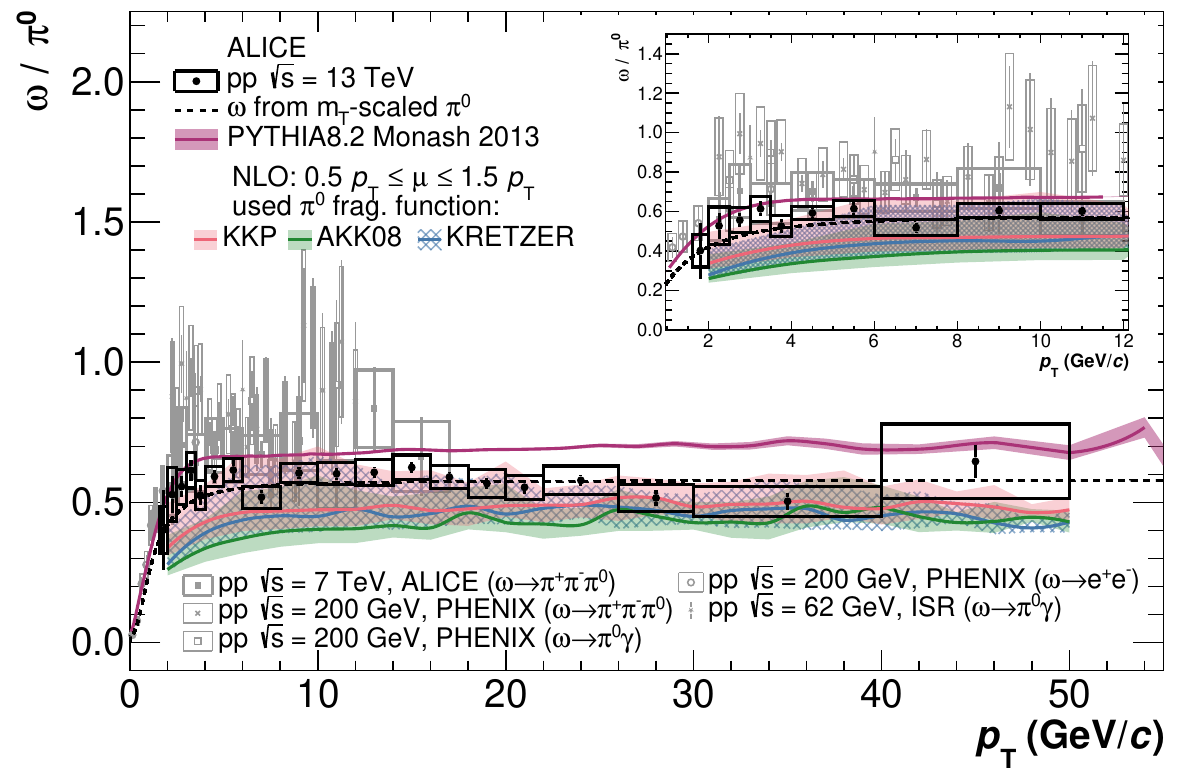}
    \caption{Ratio of $\omega/\pi^0$ production as a function of transverse momentum for pp collisions at $\sqrt{s}=\SI{13}{TeV}$. The data is compared to various measurements at lower collision energies ranging from $\sqrt{s}=62$ to \SI{7000}{GeV}~\cite{ISR,PHENIX1,PHENIX2,PHENIX3,omega7TeV}. The data is confronted with various theoretical calculations, which are given in the legend and described in the text. In addition, the ratio obtained from \mt-scaling of the measured $\pi^0$ meson cross section is shown.}
    \label{fig:omegapi0ratio}
\end{figure}

The $\omega/\pi^0$ production ratio as a function of \pt is shown in Fig.~\ref{fig:omegapi0ratio}.
It is obtained by calculating the ratio of the $\omega$ meson production cross section with respect to that of $\pi^0$ mesons, where the latter is taken from a measurement performed at the same center-of-mass energy~\cite{WAITINGPI0}.
Since both measurements were performed by the ALICE experiment using the same photon reconstruction techniques, the systematic uncertainties arising due to the material budget are largely correlated and cancel in the ratio accordingly.
A constant fit of the ratio for transverse momenta above $\SI{4}{GeV}/c$, where the ratio is mostly flat, yields $C^{\omega/\pi^0}=0.578\pm0.006~\text{(stat.)}\pm 0.013~\text{(syst.)}$, which is to our knowledge the most precise determination of this ratio to date.
As for the $\omega$ production cross section, the $\omega/\pi^0$ ratio is confronted with various theoretical calculations.
While PYTHIA8.2  using the Monash 2013 tune describes the data within the uncertainties for low \pt, the description deteriorates for $\pt\gtrsim\SI{15}{GeV}/c$, where an overestimation of the data by PYTHIA8.2  is observed for both $\omega$ and $\pi^0$ mesons.
Nonetheless, the overall description of the $\omega/\pi^0$ ratio by PYTHIA8.2  is better than for the individual cross sections, where an overestimation by up to 50\% is observed.
The ratio is furthermore confronted with NLO calculations utilizing the previously discussed model to describe $\omega$ fragmentation and three different fragmentation functions for the calculation of the $\pi^0$ production cross section.
In particular, the KKP~\cite{KKP}, AKK08~\cite{AKK08}, and KRETZER~\cite{KRETZER} functions are used to describe $\pi^0$ fragmentation.
All calculations use the CT14~\cite{CT14} parton distribution functions to describe the proton beams.
The shaded band shown in Fig.~\ref{fig:omegapi0ratio} for each calculation denotes the scale uncertainty, which was evaluated through a simultaneous variation of all scales from $0.5\pt$ to $1.5\pt$.
All three NLO calculations describe the data within the uncertainties and no significant dependence on the used fragmentation function for the $\pi^0$ reference is observed.

The $\omega/\pi^0$ production ratio measured in pp collisions at $\sqrt{s}=\SI{13}{TeV}$ is compared with data from lower collision energies at $\sqrt{s}=\num{62}$~\cite{ISR}, \num{200}~\cite{PHENIX1,PHENIX2,PHENIX3}, and \SI{7000}{GeV}~\cite{omega7TeV}, which are indicated by gray markers in Fig.~\ref{fig:omegapi0ratio}.
A significant reduction of experimental uncertainties, as well as an extension of the high-\pt reach, is achieved with respect to previous measurements.
While the ratio presented in this measurement agrees with the data at lower collision energies within the statistical and systematic uncertainties, this measurement is found to be systematically at the lower edge of the uncertainty intervals of previous measurements.

Finally, the fitted value of $C^{\omega/\pi^0}=0.578$ in the plateau region is used to test the validity of $m_{\text{\tiny T}}$-scaling~\cite{mtscaling,mtscalingwong,chargedpions}, which is an empirical scaling rule first established in measurements of identified particle spectra at ISR and RHIC~\cite{mtscalingRHIC}.
In particular, it was observed that the \pt-differential yields of most particles can be described as $E\dv*[3]{\sigma}{p}=C^h f(m_{\text{\tiny T}})$, where the spectral shape is described by the same universal function $f(m_{\text{\tiny T}})$ and the yields only differ by a constant normalization factor $C^h$ for each hadron species.
This observed scaling behavior is typically utilized to estimate hadronic background in direct photon and dielectron measurements to obtain hadron spectrum predictions in situations where no data is available.
In order to test the validity of such estimations for $\omega$ mesons, an $\omega$ cross section prediction is obtained by scaling the fit parametrization of $\pi^0$ production taken from Ref.\,\cite{WAITINGPI0} with the measured ratio $C^{\omega/\pi^0}=0.578$.
This is done following the procedure discussed in Ref.\,\cite{mtscaling} using the relation $p_{\text{\tiny T},\omega}^2+m_{0,\omega}^2=p_{\text{\tiny T},\pi^0}^2+m_{0,\pi^0}^2$ to obtain the $\pi^0$ and $\omega$ meson parametrizations, both as a function of the transverse momentum of the $\omega$ meson.
The ratio of the $\omega$ prediction obtained from $m_{\text{\tiny T}}$-scaling with respect to the $\pi^0$ parametrization is shown in Fig.~\ref{fig:omegapi0ratio} as a dashed black line.
The \mt-scaled prediction is in agreement with the data within the uncertainties, showcasing that the empirical scaling relation allows obtaining reasonable estimates of $\omega$ production over the full covered \pt~range.
This observation extends the findings from the measurement of $\omega$ meson production at $\sqrt{s}=\SI{7}{TeV}$~\cite{omega7TeV}, where the \mt-scaling prediction was found to describe the data up to the highest covered \pt of $\SI{17}{GeV}/c$.

\section{Conclusion}
\label{sec:conclusion}
The invariant differential cross section of $p+p\rightarrow\omega+X$ production at midrapidity ($|y|<0.5$) in pp collisions at $\sqrt{s}=\SI{13}{TeV}$ measured with the ALICE detector was presented.
The measurement covers an unprecedented momentum range from \num{1.6} to $\SI{50}{GeV}/c$ and is in agreement with NLO calculations over the whole covered transverse momentum range.
PYTHIA8.2 using the Monash 2013 tune is found to overestimate the data by about 50\%, as was previously reported for $\pi^0$ and $\pi^\pm$ production in the same collision system.
The $\omega/\pi^0$ ratio is found to be constant above $\SI{4}{GeV}/c$ and in agreement with NLO calculations as well as with previous measurements at lower collision energies within the uncertainties.
However, the lower uncertainties with respect to previous data reveal a slightly lower central value of the $\omega/\pi^0$ ratio than previously observed.

\newenvironment{acknowledgement}{\relax}{\relax}
\begin{acknowledgement}
\section*{Acknowledgements}
The authors would like to thank Qing Zhang for providing the NLO calculations.

The ALICE Collaboration would like to thank all its engineers and technicians for their invaluable contributions to the construction of the experiment and the CERN accelerator teams for the outstanding performance of the LHC complex.
The ALICE Collaboration gratefully acknowledges the resources and support provided by all Grid centres and the Worldwide LHC Computing Grid (WLCG) collaboration.
The ALICE Collaboration acknowledges the following funding agencies for their support in building and running the ALICE detector:
A. I. Alikhanyan National Science Laboratory (Yerevan Physics Institute) Foundation (ANSL), State Committee of Science and World Federation of Scientists (WFS), Armenia;
Austrian Academy of Sciences, Austrian Science Fund (FWF): [M 2467-N36] and Nationalstiftung f\"{u}r Forschung, Technologie und Entwicklung, Austria;
Ministry of Communications and High Technologies, National Nuclear Research Center, Azerbaijan;
Conselho Nacional de Desenvolvimento Cient\'{\i}fico e Tecnol\'{o}gico (CNPq), Financiadora de Estudos e Projetos (Finep), Funda\c{c}\~{a}o de Amparo \`{a} Pesquisa do Estado de S\~{a}o Paulo (FAPESP) and Universidade Federal do Rio Grande do Sul (UFRGS), Brazil;
Bulgarian Ministry of Education and Science, within the National Roadmap for Research Infrastructures 2020-2027 (object CERN), Bulgaria;
Ministry of Education of China (MOEC) , Ministry of Science \& Technology of China (MSTC) and National Natural Science Foundation of China (NSFC), China;
Ministry of Science and Education and Croatian Science Foundation, Croatia;
Centro de Aplicaciones Tecnol\'{o}gicas y Desarrollo Nuclear (CEADEN), Cubaenerg\'{\i}a, Cuba;
Ministry of Education, Youth and Sports of the Czech Republic, Czech Republic;
The Danish Council for Independent Research | Natural Sciences, the VILLUM FONDEN and Danish National Research Foundation (DNRF), Denmark;
Helsinki Institute of Physics (HIP), Finland;
Commissariat \`{a} l'Energie Atomique (CEA) and Institut National de Physique Nucl\'{e}aire et de Physique des Particules (IN2P3) and Centre National de la Recherche Scientifique (CNRS), France;
Bundesministerium f\"{u}r Bildung und Forschung (BMBF) and GSI Helmholtzzentrum f\"{u}r Schwerionenforschung GmbH, Germany;
General Secretariat for Research and Technology, Ministry of Education, Research and Religions, Greece;
National Research, Development and Innovation Office, Hungary;
Department of Atomic Energy Government of India (DAE), Department of Science and Technology, Government of India (DST), University Grants Commission, Government of India (UGC) and Council of Scientific and Industrial Research (CSIR), India;
National Research and Innovation Agency - BRIN, Indonesia;
Istituto Nazionale di Fisica Nucleare (INFN), Italy;
Japanese Ministry of Education, Culture, Sports, Science and Technology (MEXT) and Japan Society for the Promotion of Science (JSPS) KAKENHI, Japan;
Consejo Nacional de Ciencia (CONACYT) y Tecnolog\'{i}a, through Fondo de Cooperaci\'{o}n Internacional en Ciencia y Tecnolog\'{i}a (FONCICYT) and Direcci\'{o}n General de Asuntos del Personal Academico (DGAPA), Mexico;
Nederlandse Organisatie voor Wetenschappelijk Onderzoek (NWO), Netherlands;
The Research Council of Norway, Norway;
Pontificia Universidad Cat\'{o}lica del Per\'{u}, Peru;
Ministry of Science and Higher Education, National Science Centre and WUT ID-UB, Poland;
Korea Institute of Science and Technology Information and National Research Foundation of Korea (NRF), Republic of Korea;
Ministry of Education and Scientific Research, Institute of Atomic Physics, Ministry of Research and Innovation and Institute of Atomic Physics and Universitatea Nationala de Stiinta si Tehnologie Politehnica Bucuresti, Romania;
Ministry of Education, Science, Research and Sport of the Slovak Republic, Slovakia;
National Research Foundation of South Africa, South Africa;
Swedish Research Council (VR) and Knut \& Alice Wallenberg Foundation (KAW), Sweden;
European Organization for Nuclear Research, Switzerland;
Suranaree University of Technology (SUT), National Science and Technology Development Agency (NSTDA) and National Science, Research and Innovation Fund (NSRF via PMU-B B05F650021), Thailand;
Turkish Energy, Nuclear and Mineral Research Agency (TENMAK), Turkey;
National Academy of  Sciences of Ukraine, Ukraine;
Science and Technology Facilities Council (STFC), United Kingdom;
National Science Foundation of the United States of America (NSF) and United States Department of Energy, Office of Nuclear Physics (DOE NP), United States of America.
In addition, individual groups or members have received support from:
Czech Science Foundation (grant no. 23-07499S), Czech Republic;
FORTE project, reg.\ no.\ CZ.02.01.01/00/22\_008/0004632, Czech Republic, co-funded by the European Union, Czech Republic;
European Research Council (grant no. 950692), European Union;
ICSC - Centro Nazionale di Ricerca in High Performance Computing, Big Data and Quantum Computing, European Union - NextGenerationEU;
Academy of Finland (Center of Excellence in Quark Matter) (grant nos. 346327, 346328), Finland;
Deutsche Forschungs Gemeinschaft (DFG, German Research Foundation) ``Neutrinos and Dark Matter in Astro- and Particle Physics'' (grant no. SFB 1258), Germany.

\end{acknowledgement}

\bibliographystyle{utphys}   
\bibliography{bibliography}

\newpage
\appendix

%
%

\section{The ALICE Collaboration}
\label{app:collab}
\begin{flushleft} 
\small

S.~Acharya\,\orcidlink{0000-0002-9213-5329}\,$^{\rm 127}$, 
D.~Adamov\'{a}\,\orcidlink{0000-0002-0504-7428}\,$^{\rm 86}$, 
A.~Agarwal$^{\rm 135}$, 
G.~Aglieri Rinella\,\orcidlink{0000-0002-9611-3696}\,$^{\rm 32}$, 
L.~Aglietta\,\orcidlink{0009-0003-0763-6802}\,$^{\rm 24}$, 
M.~Agnello\,\orcidlink{0000-0002-0760-5075}\,$^{\rm 29}$, 
N.~Agrawal\,\orcidlink{0000-0003-0348-9836}\,$^{\rm 25}$, 
Z.~Ahammed\,\orcidlink{0000-0001-5241-7412}\,$^{\rm 135}$, 
S.~Ahmad\,\orcidlink{0000-0003-0497-5705}\,$^{\rm 15}$, 
S.U.~Ahn\,\orcidlink{0000-0001-8847-489X}\,$^{\rm 71}$
% ...

\section*{Affiliation Notes}

$^{\rm I}$ Deceased\\
$^{\rm II}$ Also at: Max-Planck-Institut fur Physik, Munich, Germany\\
$^{\rm III}$ Also at: Italian National Agency for New Technologies, Energy and Sustainable Economic Development (ENEA), Bologna, Italy\\
$^{\rm IV}$ Also at: Dipartimento DET del Politecnico di Torino, Turin, Italy\\
$^{\rm V}$ Also at: Department of Applied Physics, Aligarh Muslim University, Aligarh, India\\
$^{\rm VI}$ Also at: Institute of Theoretical Physics, University of Wroclaw, Poland\\
$^{\rm VII}$ Also at: Facultad de Ciencias, Universidad Nacional Aut\'{o}noma de M\'{e}xico, Mexico City, Mexico\\
$^{\rm VIII}$ Also at: An institution covered by a cooperation agreement with CERN\\

\section*{Collaboration Institutes}

$^{1}$ A.I. Alikhanyan National Science Laboratory (Yerevan Physics Institute) Foundation, Yerevan, Armenia\\
$^{2}$ AGH University of Krakow, Cracow, Poland\\
$^{3}$ Bogolyubov Institute for Theoretical Physics, National Academy of Sciences of Ukraine, Kiev, Ukraine\\
$^{4}$ Bose Institute, Department of Physics  and Centre for Astroparticle Physics and Space Science (CAPSS), Kolkata, India\\
$^{5}$ California Polytechnic State University, San Luis Obispo, California, United States\\
$^{6}$ Central China Normal University, Wuhan, China\\
$^{7}$ Centro de Aplicaciones Tecnol\'{o}gicas y Desarrollo Nuclear (CEADEN), Havana, Cuba\\
$^{8}$ Centro de Investigaci\'{o}n y de Estudios Avanzados (CINVESTAV), Mexico City and M\'{e}rida, Mexico\\
$^{9}$ Chicago State University, Chicago, Illinois, United States\\
$^{10}$ China Nuclear Data Center, China Institute of Atomic Energy, Beijing, China\\
$^{11}$ China University of Geosciences, Wuhan, China\\
$^{12}$ Chungbuk National University, Cheongju, Republic of Korea\\
$^{13}$ Comenius University Bratislava, Faculty of Mathematics, Physics and Informatics, Bratislava, Slovak Republic\\
$^{14}$ Creighton University, Omaha, Nebraska, United States\\
$^{15}$ Department of Physics, Aligarh Muslim University, Aligarh, India\\
$^{16}$ Department of Physics, Pusan National University, Pusan, Republic of Korea\\
$^{17}$ Department of Physics, Sejong University, Seoul, Republic of Korea\\
$^{18}$ Department of Physics, University of California, Berkeley, California, United States\\
$^{19}$ Department of Physics, University of Oslo, Oslo, Norway\\
$^{20}$ Department of Physics and Technology, University of Bergen, Bergen, Norway\\
$^{21}$ Dipartimento di Fisica, Universit\`{a} di Pavia, Pavia, Italy\\
$^{22}$ Dipartimento di Fisica dell'Universit\`{a} and Sezione INFN, Cagliari, Italy\\
$^{23}$ Dipartimento di Fisica dell'Universit\`{a} and Sezione INFN, Trieste, Italy\\
$^{24}$ Dipartimento di Fisica dell'Universit\`{a} and Sezione INFN, Turin, Italy\\
$^{25}$ Dipartimento di Fisica e Astronomia dell'Universit\`{a} and Sezione INFN, Bologna, Italy\\
$^{26}$ Dipartimento di Fisica e Astronomia dell'Universit\`{a} and Sezione INFN, Catania, Italy\\
$^{27}$ Dipartimento di Fisica e Astronomia dell'Universit\`{a} and Sezione INFN, Padova, Italy\\
$^{28}$ Dipartimento di Fisica `E.R.~Caianiello' dell'Universit\`{a} and Gruppo Collegato INFN, Salerno, Italy\\
$^{29}$ Dipartimento DISAT del Politecnico and Sezione INFN, Turin, Italy\\
$^{30}$ Dipartimento di Scienze MIFT, Universit\`{a} di Messina, Messina, Italy\\
$^{31}$ Dipartimento Interateneo di Fisica `M.~Merlin' and Sezione INFN, Bari, Italy\\
$^{32}$ European Organization for Nuclear Research (CERN), Geneva, Switzerland\\
$^{33}$ Faculty of Electrical Engineering, Mechanical Engineering and Naval Architecture, University of Split, Split, Croatia\\
$^{34}$ Faculty of Nuclear Sciences and Physical Engineering, Czech Technical University in Prague, Prague, Czech Republic\\
$^{35}$ Faculty of Physics, Sofia University, Sofia, Bulgaria\\
$^{36}$ Faculty of Science, P.J.~\v{S}af\'{a}rik University, Ko\v{s}ice, Slovak Republic\\
$^{37}$ Faculty of Technology, Environmental and Social Sciences, Bergen, Norway\\
$^{38}$ Frankfurt Institute for Advanced Studies, Johann Wolfgang Goethe-Universit\"{a}t Frankfurt, Frankfurt, Germany\\
$^{39}$ Fudan University, Shanghai, China\\
$^{40}$ Gangneung-Wonju National University, Gangneung, Republic of Korea\\
$^{41}$ Gauhati University, Department of Physics, Guwahati, India\\
$^{42}$ Helmholtz-Institut f\"{u}r Strahlen- und Kernphysik, Rheinische Friedrich-Wilhelms-Universit\"{a}t Bonn, Bonn, Germany\\
$^{43}$ Helsinki Institute of Physics (HIP), Helsinki, Finland\\
$^{44}$ High Energy Physics Group,  Universidad Aut\'{o}noma de Puebla, Puebla, Mexico\\
$^{45}$ Horia Hulubei National Institute of Physics and Nuclear Engineering, Bucharest, Romania\\
$^{46}$ HUN-REN Wigner Research Centre for Physics, Budapest, Hungary\\
$^{47}$ Indian Institute of Technology Bombay (IIT), Mumbai, India\\
$^{48}$ Indian Institute of Technology Indore, Indore, India\\
$^{49}$ INFN, Laboratori Nazionali di Frascati, Frascati, Italy\\
$^{50}$ INFN, Sezione di Bari, Bari, Italy\\
$^{51}$ INFN, Sezione di Bologna, Bologna, Italy\\
$^{52}$ INFN, Sezione di Cagliari, Cagliari, Italy\\
$^{53}$ INFN, Sezione di Catania, Catania, Italy\\
$^{54}$ INFN, Sezione di Padova, Padova, Italy\\
$^{55}$ INFN, Sezione di Pavia, Pavia, Italy\\
$^{56}$ INFN, Sezione di Torino, Turin, Italy\\
$^{57}$ INFN, Sezione di Trieste, Trieste, Italy\\
$^{58}$ Inha University, Incheon, Republic of Korea\\
$^{59}$ Institute for Gravitational and Subatomic Physics (GRASP), Utrecht University/Nikhef, Utrecht, Netherlands\\
$^{60}$ Institute of Experimental Physics, Slovak Academy of Sciences, Ko\v{s}ice, Slovak Republic\\
$^{61}$ Institute of Physics, Homi Bhabha National Institute, Bhubaneswar, India\\
$^{62}$ Institute of Physics of the Czech Academy of Sciences, Prague, Czech Republic\\
$^{63}$ Institute of Space Science (ISS), Bucharest, Romania\\
$^{64}$ Institut f\"{u}r Kernphysik, Johann Wolfgang Goethe-Universit\"{a}t Frankfurt, Frankfurt, Germany\\
$^{65}$ Instituto de Ciencias Nucleares, Universidad Nacional Aut\'{o}noma de M\'{e}xico, Mexico City, Mexico\\
$^{66}$ Instituto de F\'{i}sica, Universidade Federal do Rio Grande do Sul (UFRGS), Porto Alegre, Brazil\\
$^{67}$ Instituto de F\'{\i}sica, Universidad Nacional Aut\'{o}noma de M\'{e}xico, Mexico City, Mexico\\
$^{68}$ iThemba LABS, National Research Foundation, Somerset West, South Africa\\
$^{69}$ Jeonbuk National University, Jeonju, Republic of Korea\\
$^{70}$ Johann-Wolfgang-Goethe Universit\"{a}t Frankfurt Institut f\"{u}r Informatik, Fachbereich Informatik und Mathematik, Frankfurt, Germany\\
$^{71}$ Korea Institute of Science and Technology Information, Daejeon, Republic of Korea\\
$^{72}$ KTO Karatay University, Konya, Turkey\\
$^{73}$ Laboratoire de Physique Subatomique et de Cosmologie, Universit\'{e} Grenoble-Alpes, CNRS-IN2P3, Grenoble, France\\
$^{74}$ Lawrence Berkeley National Laboratory, Berkeley, California, United States\\
$^{75}$ Lund University Department of Physics, Division of Particle Physics, Lund, Sweden\\
$^{76}$ Nagasaki Institute of Applied Science, Nagasaki, Japan\\
$^{77}$ Nara Women{'}s University (NWU), Nara, Japan\\
$^{78}$ National and Kapodistrian University of Athens, School of Science, Department of Physics , Athens, Greece\\
$^{79}$ National Centre for Nuclear Research, Warsaw, Poland\\
$^{80}$ National Institute of Science Education and Research, Homi Bhabha National Institute, Jatni, India\\
$^{81}$ National Nuclear Research Center, Baku, Azerbaijan\\
$^{82}$ National Research and Innovation Agency - BRIN, Jakarta, Indonesia\\
$^{83}$ Niels Bohr Institute, University of Copenhagen, Copenhagen, Denmark\\
$^{84}$ Nikhef, National institute for subatomic physics, Amsterdam, Netherlands\\
$^{85}$ Nuclear Physics Group, STFC Daresbury Laboratory, Daresbury, United Kingdom\\
$^{86}$ Nuclear Physics Institute of the Czech Academy of Sciences, Husinec-\v{R}e\v{z}, Czech Republic\\
$^{87}$ Oak Ridge National Laboratory, Oak Ridge, Tennessee, United States\\
$^{88}$ Ohio State University, Columbus, Ohio, United States\\
$^{89}$ Physics department, Faculty of science, University of Zagreb, Zagreb, Croatia\\
$^{90}$ Physics Department, Panjab University, Chandigarh, India\\
$^{91}$ Physics Department, University of Jammu, Jammu, India\\
$^{92}$ Physics Program and International Institute for Sustainability with Knotted Chiral Meta Matter (WPI-SKCM$^{2}$), Hiroshima University, Hiroshima, Japan\\
$^{93}$ Physikalisches Institut, Eberhard-Karls-Universit\"{a}t T\"{u}bingen, T\"{u}bingen, Germany\\
$^{94}$ Physikalisches Institut, Ruprecht-Karls-Universit\"{a}t Heidelberg, Heidelberg, Germany\\
$^{95}$ Physik Department, Technische Universit\"{a}t M\"{u}nchen, Munich, Germany\\
$^{96}$ Politecnico di Bari and Sezione INFN, Bari, Italy\\
$^{97}$ Research Division and ExtreMe Matter Institute EMMI, GSI Helmholtzzentrum f\"ur Schwerionenforschung GmbH, Darmstadt, Germany\\
$^{98}$ Saga University, Saga, Japan\\
$^{99}$ Saha Institute of Nuclear Physics, Homi Bhabha National Institute, Kolkata, India\\
$^{100}$ School of Physics and Astronomy, University of Birmingham, Birmingham, United Kingdom\\
$^{101}$ Secci\'{o}n F\'{\i}sica, Departamento de Ciencias, Pontificia Universidad Cat\'{o}lica del Per\'{u}, Lima, Peru\\
$^{102}$ Stefan Meyer Institut f\"{u}r Subatomare Physik (SMI), Vienna, Austria\\
$^{103}$ SUBATECH, IMT Atlantique, Nantes Universit\'{e}, CNRS-IN2P3, Nantes, France\\
$^{104}$ Sungkyunkwan University, Suwon City, Republic of Korea\\
$^{105}$ Suranaree University of Technology, Nakhon Ratchasima, Thailand\\
$^{106}$ Technical University of Ko\v{s}ice, Ko\v{s}ice, Slovak Republic\\
$^{107}$ The Henryk Niewodniczanski Institute of Nuclear Physics, Polish Academy of Sciences, Cracow, Poland\\
$^{108}$ The University of Texas at Austin, Austin, Texas, United States\\
$^{109}$ Universidad Aut\'{o}noma de Sinaloa, Culiac\'{a}n, Mexico\\
$^{110}$ Universidade de S\~{a}o Paulo (USP), S\~{a}o Paulo, Brazil\\
$^{111}$ Universidade Estadual de Campinas (UNICAMP), Campinas, Brazil\\
$^{112}$ Universidade Federal do ABC, Santo Andre, Brazil\\
$^{113}$ Universitatea Nationala de Stiinta si Tehnologie Politehnica Bucuresti, Bucharest, Romania\\
$^{114}$ University of Cape Town, Cape Town, South Africa\\
$^{115}$ University of Derby, Derby, United Kingdom\\
$^{116}$ University of Houston, Houston, Texas, United States\\
$^{117}$ University of Jyv\"{a}skyl\"{a}, Jyv\"{a}skyl\"{a}, Finland\\
$^{118}$ University of Kansas, Lawrence, Kansas, United States\\
$^{119}$ University of Liverpool, Liverpool, United Kingdom\\
$^{120}$ University of Science and Technology of China, Hefei, China\\
$^{121}$ University of South-Eastern Norway, Kongsberg, Norway\\
$^{122}$ University of Tennessee, Knoxville, Tennessee, United States\\
$^{123}$ University of the Witwatersrand, Johannesburg, South Africa\\
$^{124}$ University of Tokyo, Tokyo, Japan\\
$^{125}$ University of Tsukuba, Tsukuba, Japan\\
$^{126}$ Universit\"{a}t M\"{u}nster, Institut f\"{u}r Kernphysik, M\"{u}nster, Germany\\
$^{127}$ Universit\'{e} Clermont Auvergne, CNRS/IN2P3, LPC, Clermont-Ferrand, France\\
$^{128}$ Universit\'{e} de Lyon, CNRS/IN2P3, Institut de Physique des 2 Infinis de Lyon, Lyon, France\\
$^{129}$ Universit\'{e} de Strasbourg, CNRS, IPHC UMR 7178, F-67000 Strasbourg, France, Strasbourg, France\\
$^{130}$ Universit\'{e} Paris-Saclay, Centre d'Etudes de Saclay (CEA), IRFU, D\'{e}partment de Physique Nucl\'{e}aire (DPhN), Saclay, France\\
$^{131}$ Universit\'{e}  Paris-Saclay, CNRS/IN2P3, IJCLab, Orsay, France\\
$^{132}$ Universit\`{a} degli Studi di Foggia, Foggia, Italy\\
$^{133}$ Universit\`{a} del Piemonte Orientale, Vercelli, Italy\\
$^{134}$ Universit\`{a} di Brescia, Brescia, Italy\\
$^{135}$ Variable Energy Cyclotron Centre, Homi Bhabha National Institute, Kolkata, India\\
$^{136}$ Warsaw University of Technology, Warsaw, Poland\\
$^{137}$ Wayne State University, Detroit, Michigan, United States\\
$^{138}$ Yale University, New Haven, Connecticut, United States\\
$^{139}$ Yonsei University, Seoul, Republic of Korea\\
$^{140}$  Zentrum  f\"{u}r Technologie und Transfer (ZTT), Worms, Germany\\
$^{141}$ Affiliated with an institute covered by a cooperation agreement with CERN\\
$^{142}$ Affiliated with an international laboratory covered by a cooperation agreement with CERN.\\

\end{flushleft} 

\end{document}